\documentclass{aastex}
\usepackage{spr-astr-addons}

\RequirePackage{color}

\def\simless{\mathbin{\lower 3pt\hbox
     {$\rlap{\raise 5pt\hbox{$\char'074$}}\mathchar"7218$}}}   
\def\simmore{\mathbin{\lower 3pt\hbox
     {$\rlap{\raise 5pt\hbox{$\char'076$}}\mathchar"7218$}}}   
\def\msun{{\rm M}_\odot}                                       
\newcommand{\ha}  {H$\alpha$}
\newcommand{\ew}  {EW(H$\alpha$)}
\newcommand{\bex}  {BeXB}

\begin{document}

\title{Be/X-ray Binaries}
\shorttitle{Be/X-ray Binaries}
\shortauthors{P. Reig}

\author{Pablo Reig\altaffilmark{1,2}}
\altaffiltext{1}{IESL, Foundation for Reseach and Technology-Hellas, 71110,  Heraklion, Greece.}
\altaffiltext{2}{Physics Department, University of Crete, 71003, Heraklion, Greece.}

\begin{abstract}

The interest in X/$\gamma$-ray Astronomy has grown enormously in the last
decades thanks to the ability to send X-ray space missions above the
Earth's atmosphere. There are more than half a million X-ray sources
detected and over a hundred missions (past and currently operational)
devoted to the study of cosmic X/$\gamma$ rays.  With the improved
sensibilities of the currently active missions new detections occur almost
on a daily basis.   Among these, neutron-star X-ray binaries form an
important group because they are among the brightest extra-solar objects in
the sky and are characterized by dramatic variability in brightness on
timescales ranging from  milliseconds to months and years. Their main
source of power is the gravitational energy released by matter accreted
from a companion star and falling onto the neutron star in a relatively close
binary system.

Neutron-star X-ray binaries divide into high-mass and low-mass systems
according to whether the mass of the donor star is above $\sim$8 or below
$\sim$2 $\msun$, respectively. Massive X-ray binaries divide further into
supergiant X-ray binaries and Be/X-ray binaries depending on the
evolutionary status of the optical companion.  Virtually all Be/X-ray
binaries show X-ray pulsations. Therefore,  these systems
can be used as unique natural laboratories to investigate the properties
of matter  under extreme conditions of gravity and magnetic field.

The purpose of this work is to review the observational properties of
Be/X-ray binaries.  The open questions in Be/X-ray binaries include those
related to the Be star companion, that is, the so-called "Be phenomenon",
such as, timescales associated to the formation and dissipation of the
equatorial disc, mass-ejection mechanisms, V/R variability, and rotation
rates;  those related to the neutron star, such as, mass determination,
accretion physics, and spin period evolution; but also, those that result
from the interaction of the two constituents, such as,  disc truncation and
mass transfer. Until recently, it was thought that the Be stars' disc was
not significantly affected by the neutron star. In this review, I present
the observational evidence accumulated in recent years on the interaction
between the circumstellar disc and the compact companion. The most obvious
effect is the tidal truncation of the disc. As a result, the equatorial
discs in Be/X-ray binaries are smaller and denser than those around
isolated Be stars. 

\end{abstract}


\section{Definition and classification of X-ray binaries}

\begin{figure}[t]
\includegraphics[width=0.9\linewidth]{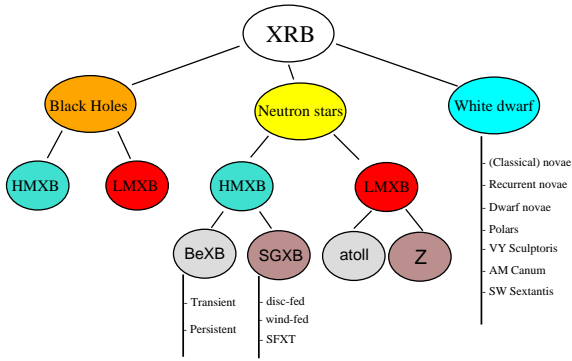} 
\caption{Classification of X-ray binaries.}
\label{xrb_class}
\end{figure}

 In very general terms, one can simply define X-ray binaries as systems
that consist of a compact object orbiting an optical companion. They are
"close" binary systems because there exists a transfer of mass from the
optical component to the compact object. By "optical companion" it is
understood that nuclear burning is still taking place in its interior.
Figure~\ref{xrb_class} shows a tree-diagram depicting all the different
subsystems that comprise the generic group of X-ray binaries. 

In referring to the two components in an X-ray binary one should be careful
and learn which is the subject of investigation as the same name can be
used to mean different components. In massive X-ray binaries, the most
massive star is normally termed {\em primary} whereas the less massive one
is called  {\em secondary}. In low-mass systems, the term {\em primary}
refers to the neutron star while the word {\em secondary} is reserved for
the late-type companion.  Other names also used are "optical" or "donor"
for the larger star and "compact", "gainer" or "accreting" for the denser
companion. 

They admit several classification schemes depending upon whether the
emphasis is put on the type of the compact companion or the physical
properties of the optical star. X-ray binaries divide up into black-hole
systems, neutron star X-ray binaries or cataclysmic variables (if the
compact object is a white dwarf).  Nevertheless, the term "X-ray binaries"
is  normally reserved to designate binaries with neutron stars.

\subsection{High-mass X-ray binaries}
\label{intro_hmxb}

Neutron-star X-ray binaries are divided up into high-mass (HMXBs) and
low-mass (LMXBs) X-ray  binaries  depending on the spectral type of the 
mass  donor,  as  this  feature  determines  the  mode  of transferring  
mass  to  the  compact   object  and  the   environment surrounding the
X-ray source. HMXBs contain early-type (O or B) companions, while the
spectral type of the optical star in LMXBs is later than A. HMXBs are
strong emitters of X-ray radiation. Sometimes they appear as the brightest
objects of the X-ray sky. The high-energy radiation is produced as the
result of accretion of matter from the optical companion onto the neutron
star. The term accretion refers to the gradual accumulation or deposition
of matter onto the surface of an object under the influence of gravity. If
the accreting object is a neutron star (or black hole), then matter falls
down onto an enormous well of gravitational potential and is accelerated to
extremely high velocities. When the matter reaches the surface of the
neutron star, it is rapidly decelerated and the free-fall kinetic energy
radiated away as heat which is available to power the X-ray source.

The luminosity class serves to subdivide HMXBs into Be/X-ray binaries
(\bex), when the optical star is a dwarf, subgiant or giant OBe star
(luminosity class III, IV or V) and supergiant X-ray binaries (SGXBs), if
they contain a luminosity class I-II star. 

In SGXBs, the optical star emits a  substantial  stellar  wind, removing 
between  $10^{-6}-10^{-8}$  M$_{\odot}$  yr$^{-1}$  with a terminal 
velocity up to 2000 km s$^{-1}$.  A neutron star in a relatively  close
orbit will capture some fraction of this wind, sufficient to power a bright
X-ray source. If  mass  transfer occurs via Roche lobe overflow, then the
X-ray emission is highly enhanced and an accretion disc is formed around
the neutron star. At present, there is known only one disc-fed SGXB in the
Galaxy (Cen X-3) and three in total (SMC X-1 and LMC X-4), while there are
about a few tens of wind-fed SGXBs. Because of their brightness and
persistent X-ray emission, SGXBs were the first to be discovered.  They
were initially thought to represent the dominating population of HMXBs,
whereas \bex\ were considered atypical cases. Hence, the name {\em
classical} or {\em standard} was given to SGXBs.

In \bex, the optical companion is a Be star. Be stars are non-supergiant
fast-rotating B-type and luminosity class III-V stars which at some point
of their lives have shown spectral lines in emission, hence the qualifier
"$e$" in their spectral types \citep{port03, balo00, slet88}. The best
studied lines are those of hydrogen (Balmer and Paschen series) but Be
stars may also show He, Fe in emission \citep[see e.g][]{hanu96}. They also
show an amount of IR radiation than is larger than that expected from an
absorption-line B star of the same spectral type. This extra
long-wavelength emission is known as infrared excess. 

The origin of the emission lines and infrared excess in \bex\ is attributed
to an equatorial disc, fed from material expelled from the rapidly rotating
Be star in a manner that it is not yet understood \citep{port03}. During
periastron, the neutron star passes close to this disc, sometimes may even
go through it causing major disruption. A large flow of matter is then
accreted onto the compact object. The conversion of the
kinetic energy of the in-falling matter into radiation powers the X rays.
\bex\ have large orbital periods and by definition they are non-supergiant
systems. Hence, the Be star is well within the Roche lobe. However,
transient Roche lobe overflow may occur during periastron passage in
systems with large eccentric orbits or during giant X-ray outbursts when a
large fraction of the Be star's disc is believed to be accreted.

Most \bex\ are transient systems and present moderately eccentric orbits
($e\simmore0.3$), although persistent sources also exist \citep{reig99}.
Persistent \bex\ differ from transient \bex\ in that  they display much less
X-ray variability (no large outbursts are detected), lower luminosity ($L_x
\simless 10 ^{35}$ erg s$^{-1}$), contain slowly rotating neutron
stars  ($P_{\rm spin} > 200$ s) and reside in wider orbit systems ($P_{\rm
orb} > 200$ d).

Table \ref{bexlist} lists the confirmed \bex. This table includes only
those systems with identified optical counterparts and whose X-ray and
optical/IR variability has been shown to be typical of \bex\ (see
Sects.~\ref{optir} and \ref{xrayvar}). Table \ref{stat} gives the number
of various types of X-ray binaries. There are more than 300 bright X-ray
sources with  fluxes well above $10^{-10}$ erg cm$^{-2}$ s$^{-1}$ in the
energy range 1--10 keV  \citep{liu06,liu07,bird10}. The distribution of
these sources shows a clear concentration towards the Galactic center and
also towards the Galactic plane, indicating that the majority belong indeed
to our Galaxy.

\begin{deluxetable}{lllllccccccc}
\tabletypesize{\scriptsize}
\rotate
\label{bexlist} 
\tablecaption{List of galactic \bex. Only systems with known optical
counterparts and well-established optical/X-ray behaviour are included}
\tablewidth{0pt}
\tablehead{
\colhead{X-ray name}	&\colhead{Optical Ctp.}&\colhead{RA (J2000)}	&\colhead{Dec (J2000)}	
&\colhead{Spec. type}	&\colhead{V}	&\colhead{J}		&\colhead{E(B-V)}
&\colhead{$P_{\rm spin}$ (s)}	  &\colhead{$P_{\rm orb}$ (d)}    &\colhead{e}      
&\colhead{d} (kpc)} 
\startdata
4U 0115+634	&V635 Cas	&01 18 31.90	&+63 44 24.0	 &B0.2 Ve 	&14.8-15.5&10.8-12.3	&1.55	 &3.6	  &24.3    &0.34    &8       \\
IGR J01363+6610	&--		&01 36 18.00	&+66 10 36.0	 &B1 IV-Ve	&13.3	 &10	  	&1.61	 &--	  &--	   &--      &2       \\
RX J0146.9+6121 &LS I +61 235	&01 47 00.17	&+61 21 23.7	 &B1Ve    	&11.2	 &10	  	&0.93	 &1412    &330     &--      &2.3     \\
IGR J01583+6713	&--		&01 58 18.20	&+67 13 26.0	 &B2IVe   	&14.4	 &11.5    	&1.46	 &19692   &--	   &--      &4       \\
RX J0240.4+6112	&LS I +61 303	&02 40 31.67	&+61 13 45.6	 &B0.5 Ve 	&10.7	 &8.8	  	&0.75	 &--	  &26.45   &0.54    &3.1     \\
V 0331+530	&BQ Cam		&03 34 59.89	&+53 10 23.6	 &O8-9 Ve 	&15.1-15.8&11.4-12.2	&1.9	 &4.4	  &34.3    &0.3     &7       \\
4U 0352+309 (X Per)&HD 24534	&03 55 23.08	&+31 02 45.0	 &O9.5IIIe-B0Ve &6.1-6.8 &5.7-6.5 	&0.4	 &837	  &250     &0.11    &1.3     \\
RX J0440.9+4431	&LS V +44 17	&04 40 59.32	&+44 31 49.3	 &B0.2Ve  	&10.8	 &9.2	  	&0.65	 &203	  &--	   &--      &3.3     \\
1A 0535+262	&V725 Tau	&05 38 54.57	&+26 18 56.8	 &B0 IIIe 	&8.9-9.6 &7.7-8.5 	&0.75	 &105	  &111     &0.47    &2.4     \\
IGR J06074+2205	&--		&06 07 24.00	&+22 05 00.0	 &B0.5V   	&12.3	 &10.5	  	&1.88	 &--	  &--	   &--      &5       \\
XTE J0658-073	&--		&06 58 42.00	&--07 11 00.0	 &O9.7 Ve 	&12.1    &9.7	  	&1.19	 &160.4   &--	   &--      &3.9     \\
4U 0726-260	&V441 Pup	&07 28 53.60	&--26 06 29.0	 &O8-9Ve  	&11.6	 &10.4    	&0.73	 &103.2   &34.5    &--      &6       \\
RX J0812.4-3114	&LS 992		&08 12 28.84	&--31 14 52.2	 &B0.2 III-IVe  &12.4    &11.2-12.0	&0.65	 &31.89   &80	   &--      &9       \\
GS 0834-430	&--		&08 35 55.00	&--43 11 06.0	 &B0-2 III-Ve	&20.4	 &13.3	  	&4.0	 &12.3    &105.8   &0.12    &5       \\
GRO J1008-57	&--		&10 09 46.00	&--58 17 30.0	 &O9e-B1e 	&15.3    &10.9    	&1.9	 &93.5    &247.5   &0.66    &2       \\
RX J1037.5-5647	&LS 1698	&10 37 35.50	&--56 48 11.0	 &B0 III-Ve	&11.3	 &--	  	&0.75	 &862	  &--	   &--      &5       \\
1A 1118-615	&Hen 3-640	&11 20 57.21	&--61 55 00.3	 &O9.5 III-Ve	&12.1	 &9.6	  	&0.92	 &405	  &--	   &--      &5       \\
4U 1145-619	&V801 Cen	&11 48 00.02	&--62 12 24.9	 &B1 Vne  	&9.3	 &8.7	  	&0.29	 &292.4   &187.5   &$>$0.5  &3.1     \\
4U 1258-613	&GX 304-1	&13 01 17.20	&--61 36 07.0	 &B2 Vne  	&13.5	 &9.8	  	&2.0	 &272	  &132.5   &$>$0.5  &2.4     \\
2S 1417-624	&--		&14 21 12.80	&--62 41 54.0	 &B1 Ve   	&17.2	 &13.3    	&2.0	 &17.6    &42.12   &0.45    &10      \\
GS 1843+00	&--		&18 45 36.90	&+00 51 48.2	 &B0-2 IV-Ve	&20.9	 &13.7	  	&2.5	 &29.5    &--	   &--      &$>$10   \\
XTE J1946+274	&--		&19 45 39.30	&+27 21 55.0	 &B0-1 IV-Ve	&16.8	 &12.5	  	&2.0	 &15.8    &169.2   &0.33    &8-10    \\
KS 1947+300	&--		&19 49 30.50	&+30 12 24.0	 &B0 Ve   	&14.2	 &11.7    	&1.09	 &18.76   &40.4    &0.03    &9.5     \\
EXO 2030+375	&V2246 Cyg	&20 32 15.20	&+37 38 15.0	 &B0e	  	&19.7	 &12.1    	&3.8	 &41.8    &46.03   &0.41    &5       \\
GRO J2058+42	&--		&20 59 00.00	&+41 43 00.0	 &O9.5-B0 IV-Ve &14.9    &11.7    	&1.38	 &192	  &110     &--      &9       \\
SAX J2103.5+4545&--		&21 03 35.71	&+45 45 05.5	 &B0 Ve   	&14.2	 &11.8    	&1.35	 &358.6   &12.67   &0.4     &6.8     \\
4U 2135+57	&Cep X-4	&21 39 30.72	&+56 59 10.0	 &B1-B2 Ve	&14.2	 &11.8    	&1.3	 &66.3    &--	   &--      &3.8     \\
SAX J2239.3+6116&--		&22 39 20.90	&+61 16 26.8	 &B0-2 III-Ve	&15.1	 &11.5    	&1.4	 &1247    &262.6   &--      &4.4     \\
\enddata
\end{deluxetable}

\begin{table*} 
\begin{center}
\caption{Statistics on HMXBs in the Milky Way}
\label{stat} 
\begin{tabular}{@{}lc@{}} 
\tableline 
Number of neutron-star X-ray binaries\tablenotemark{1}		&327	\\
Number of suspected HMXB\tablenotemark{1}			&131	\\
Number of suspected \bex\tablenotemark{2}			&63	\\
Number of {\em confirmed} \bex\tablenotemark{3}			&28	\\
\tableline
\end{tabular}
\tablenotetext{1}{Sources in the \citet{liu07} and \citet{liu06} catalogs plus 
update from the 4th IBIS/ISGRI soft gamma-ray survey catalog \citep{bird10}}
\tablenotetext{2}{Sources in the updated on-line version of the \citet{ragu05}
catalog (http://xray.sai.msu.ru/$^\sim$raguzova/BeXcat/)}
\tablenotetext{3}{Systems with known optical counterpart {\em and} verified X-ray
behaviour (from Table~\ref{bexlist})}
\end{center}
\end{table*}

Until the advent of the INTEGRAL mission in 2002, the number of \bex\ was
growing fast, while the number of SGXBs had stabilised. During the 80's and
90's the rate of new discoveries was approximately four \bex\ for one
SGXB. The reason lies in the different origin of the accreted mass 
(stellar wind  versus Be star's envelope), leading to persistent emission
in SGXBs and transient emission in \bex. Since SGXBs are persistent
sources, new discoveries come from the improvement of the sensitivity of
the X-ray detectors on board space missions. \bex\ benefited from the
technical advances too but the discovery of new systems is also related to
their triggering mechanism being activated. In addition, the evolutionary
time scales involved imply that SGXBs are less numerous than \bex. The
accretion-powered phase is relatively short for OB supergiant systems (of
the order of 10000 years).

\begin{figure}[t]
\begin{center}
\includegraphics[width=0.9\linewidth]{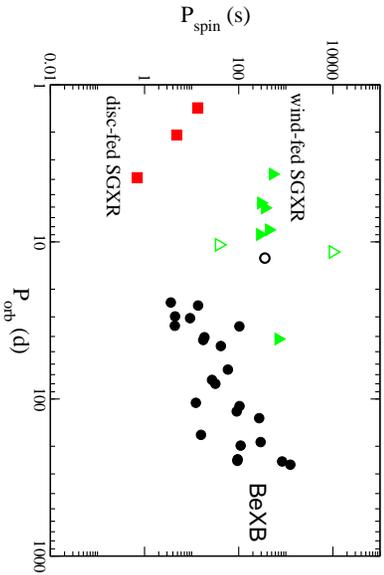}
\caption{$P_{\rm orb}-P_{\rm spin}$ diagram.
The open triangles correspond to 2S 0114+65 and OAO 1657-41 and the open
circle to SAX J2103.5+4545.}
\label{pspo}
\end{center}
\end{figure}

The different types of HMXBs occupy well-defined positions in the spin
period versus orbital period diagram \citep{corb86}, which reflects the
different types of mass transfer. SGXBs exhibit no correlation at all or an
anticorrelation, while \bex\ show a positive correlation in this diagram.
Figure~\ref{pspo} displays an updated version of Corbet's diagram. Disc-fed
SGXBs (squares) show short orbital periods and short spin periods and
display an anticorrelation in the $P_{\rm orb}-P_{\rm spin}$ diagram. The
small orbital separation and evolved companions make Roche lobe overflow
the most likely mass transfer mechanism. Wind-fed SGXBs (triangles) show
long spin periods and short orbital periods, occupying a more or less flat
region in the  $P_{\rm orb}-P_{\rm spin}$ diagram. Two systems (open
triangles) prevent the region from being a horizontal line:  OAO1657-41,
which might be making a transition to the disc-fed SGXB and 2S 0114+65, for
which the association of the $10^4$ s pulsations with the spin period of
the neutron star remains controversial \citep{koenigsberger06}. The spin
and orbital periods of \bex\ (filled circles) exhibit a clear correlation.
The open circle in Fig.~\ref{pspo} represents SAX J2103.5+4545, a peculiar
system whose X-ray properties are similar to those of wind-fed systems but
whose optical/IR emission resemble that of \bex\ \citep{reig10a}.

The observed correlation in Be/X systems is explained in terms of the
equilibrium period, defined as the period at which the outer edge of the
magnetosphere rotates with the Keplerian velocity
\citep{davi73,stel86,wate89}. If the the neutron star (and hence the
magnetosphere) rotates faster that the equilibrium period, then matter is
spun away by the propeller mechanism; only when the spin period is larger
than the equilibrium period can matter be accreted on to the neutron star
surface. This results in angular momentum transfer to the neutron star,
increasing its rotation velocity (decreasing the spin period). The
equilibrium period depends mainly on the mass flux (or accretion rate)
because it determines the size of the magnetosphere which is assumed to
corotate with the neutron star. In turn, the accretion rate depends on the
separation of the two components of the binary systems, hence on the
orbital period. 

The compact object in all confirmed \bex\ (Table~\ref{bexlist}) is a
neutron star. In fact, many times the neutron star is taken as a defining
property of \bex. It is common to read in the literature that a \bex\
consists of a neutron star and a Be star. However, there seems to be no
apparent mechanism that would prevent the formation of Be stars with black
holes (BH) or white dwarfs (WD). Surprisingly, not a single \bex\ is known
to host a black-hole companion in our Galaxy, whereas the interpretation of
$\gamma$ Cas as a Be+WD system still remains very controversial. 

Two ideas have been put forward to explain this apparent lack of Be/BH
binaries. \citet{zhan04a} extended the application of the viscous decretion
disc model of \citet{okaz01} to compact companions of arbitrary mass  and
showed that the most effective Be disc truncation would occur in relatively
narrow systems. Using the population synthesis results of \citet{pods03},
which state that binary black holes are most likely to be born in systems
with narrow orbits ($P_{\rm orb}\simless 30$ days), the reason for the lack
of these systems can be attributed to the difficulty to detecting them.
Be/BH binaries are expected to be X-ray transients with very long quiescent
states. In contrast, \citet{belc09} showed that Be/BH binaries do not
necessarily have narrow orbits. These authors argued that the predicted
ratio of Be/NS binaries to Be/BH binaries ($F_{\rm NS-to-BH}\sim 30-50$)
in our Galaxy, based on current population synthesis models and evolutionary
scenarios, is consistent with the observations: there are 60 known Be/NS
binaries, hence one would expect 0--2 Be/BH binaries, consistent with the
observed galactic sample.  Both \citet{zhan04a} and \citet{belc09} agree in
that Be/BH binaries should exist in the Galaxy.

\subsection{Other types of high-mass X-ray binaries}

The traditional picture of two classes of HMXBs, namely SGXB --- subdivided
into low-luminosity or wind-fed systems and high-luminosity or
disc-fed systems --- and \bex\ --- transient and persistent --- is giving way
to a more complex situation where newly discovered systems may not fit in
these categories.

\subsubsection{Low eccentricity \bex}
\label{low-e}

There is a group of so far five \bex\ (X Per, GS 0834--430, KS 1947+300,
XTE J1543--568, and 2S 1553-542\footnote{The optical counterparts of XTE
J1543--568 and 2S 1553-542 are not known, hence they do not appear in
Table~\ref{bexlist}.}), characterised by $P_{\rm orb} \simmore$ 30 d and
very low eccentricity ($e\simless 0.2$).  Their low eccentricity requires
that the neutron star received a much lower kick velocity at birth than
previously assumed by current evolutionary models \citep{pfah02}.

Most popular models for neutron star kicks involve a momentum impulse
delivered around the time of the core collapse that produced the neutron
star. These models assume that neutron stars are born with speeds in excess
of 100-200 km s$^{-1}$. Such velocities imply a probability close to zero
that the post-supernova eccentricity is less than 0.2. Hence, the discovery
of these low-eccentricity \bex\ with wide orbits was unexpected. Tidal
circularizacion is ruled out since this mechanism requires that the star
almost fills its Roche lobe and that there be an effective mechanism for
damping the tide. Tidal torques should have little effect on the orbit of a
HMXB with  $P_{\rm orb}=10$ days, as long as the secondary is not too
evolved and the eccentricity is not so large that the tidal interaction is
enhanced dramatically at periastron (as it is the case of typical \bex).

\citet{pfah02} developed a phenomenological model that simultaneously
accounts for the long-period ($\simmore$30 d), low-eccentricity
($\simless$0.2) HMXBs and which does not violate any previous notions
regarding the numbers and kinematics of other neutron star populations.
They propose that a neutron star receives a relatively small kick
($\simless$50 km s$^{-1}$) if the progenitor, i.e., the core of the
initially more massive star in the binary, rotates rapidly. This condition
may be met when the progenitor star experienced case B$_e$ or C$_e$ mass
transfer in a binary system\footnote{In case B, mass transfer occurs during
the shell hydrogen-burning phase, but prior to central helium ignition,
while case C evolution begins after helium has been depleted in the core.
Cases B and C are naturally divided into an early case (B$_e$ or C$_e$),
where the envelope of the primary is mostly radiative, and a late case
(B$_l$ or C$_l$), where the primary has a deep convective envelope.}. If
the hydrogen-exhausted core of an initially rapidly rotating massive star
is exposed following case Be or Ce mass transfer in a binary, then the core
is also likely to be a rapid rotator.

\subsubsection{Obscured sources \& Supergiant fast X-ray transients}

Since the launch of {\it INTEGRAL} in October 2002, the situation is also
changing among the SGXB group. INTEGRAL has unveiled a population of 
highly obscured HMXBs with supergiant companions and a new type of source
displaying outbursts which are significantly shorter than typical for \bex\ and
which are characterized by bright flares with a duration of a few hours and
peak luminosities of $10^{36}-10^{37}$ erg s$^{-1}$. These new systems have
been termed as Supergiant fast X-ray Transients 
\citep[SFXTs,][]{smit06,negu06,walt07,negu08}. 

Both obscured HMXBs and SFXTs display X-ray and IR spectra typical of
SGXBs. In some cases, the X-ray sources are pulsed and orbital parameters
typical of persistent SGXBs have been found
\citep[e.g.,][]{boda06,zuri06}.   The heavily-absorbed sources had not been
detected by previous missions due to high absorption, which renders their
spectra very hard. The current understanding is that the entire binary
system is surrounded by a dense and absorbing circumstellar material
envelope or cocoon, made of cold gas and/or dust \citep{chaty08}. 

SFXTs differ from SGXBs because they are only detected sporadically, during
very brief outbursts \citep{romano10}. A promising model to explain SFXTs
invokes highly structured (clumpy) stellar winds. The outburst occurs as a
result of the accretion of one of the clumps of dense matter from the wind.
An alternative model ("Be-type" model) assumes a very elliptical orbit for
the binary. In this model outbursts are triggered when the compact object
travels through its periastron. Other possibilities imply that SFXTs
contain strongly magnetised neutron stars. The outbursts result from the
overcoming of centrifugal and magnetic barriers \citep[see][and references
therein]{grebenev10}.

\begin{figure}[t]
\begin{center}
\includegraphics[width=0.9\linewidth]{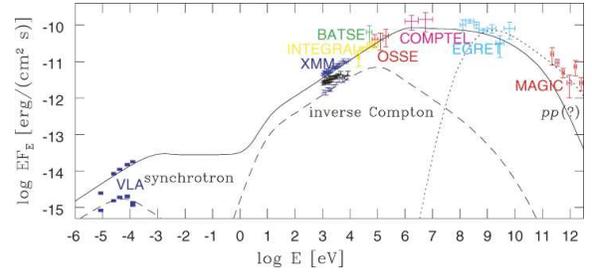} 
\caption{Spectral energy distribution of LSI +61 303. From
\citet{chernyakova06}}.
\label{grb_sed}
\end{center}
\end{figure}

\subsubsection{$\gamma$-ray binaries}
\label{grb}

 $\gamma$-ray binaries are HMXBs that emit most of
their radiative output in the MeV-TeV range.  Currently, only four HMXBs are
well-established members of this group of high-energy sources: LS I
+61 303, LS 5039, PSR B1259-63, and Cygnus X-3, while other two are firm
candidates: Cygnus X-1, and HESS J0632+057\footnote{The
Be star HD 215227, likely  counterpart of the gamma-ray source AGL
J2241+4454, has been suggested as a new candidate \citep{williams10}.}.
The optical counterpart is either a luminous O-type star (Cygnus X-1, LS
5039), a Be star (LS I +61 303, PSR B1259-63, HESS J0632+057) or a likely
Wolf-Rayet star in Cygnus X-3. In addition to the variable $\gamma$-ray
emission these systems share in common a resolved radio counterpart with a
jet or jet-like structure, multi-wavelength orbital modulation and 
spectral energy distribution (Fig.~\ref{grb_sed}). The wide range of 
orbital parameters \citep{paredes08} and the non-unique nature of the
compact companion (unknown in most systems but with a confirmed neutron
star in PSR B1259-63 and a fairly secure black hole in Cyg X-1) represent a
challenge for the theoretical modelling of these systems.  Two alternative
scenarios may explain the variable $\gamma$-ray emission: the microquasar
or accretion-powered scenario and the pulsar wind scenario.

All confirmed $\gamma$-ray binaries show a jet or jet-like radio structure,
which would indicate the presence of relativistic particles. In black-hole
binaries the radio jet can account for their broad-band spectrum, from
radio to X-rays \citep{markoff03} as well as for the origin of most of the
timing variability \citep{kylafis08}. Therefore, it is reasonable to think
that the jet may also be the origin of the very high-energy $\gamma$-rays.
In the context of relativistic jets, the most efficient gamma-ray
mechanism would be inverse Compton scattering, by which relativistic
particles collide with stellar and/or synchrotron photons and boost their
energies to the VHE range.  Two flavors of the microquasar model can be
found in the literature depending on whether hadronic or leptonic jet
matter dominates the emission at such an energy range. Among leptonic jet
models, there are inverse Compton jet emission models in which X-rays and
$\gamma$-rays result from synchrotron self-Compton processes
\citep{atoyan99}, or in which the seed photons  come from external sources,
i.e., companion star and/or accretion disk photons
\citep{kaufman02,georganopoulos02}. In the hadronic scenario, the gamma-ray
emission arises from the decay of neutral pions created in the inelastic
collisions between relativistic protons in the jet and either the ions in
the stellar wind of the massive companion star \citep{romero03} or nearby
high-density regions (i.e. molecular clouds) \citep{bosch05}.

Alternatively, relativistic particles can be injected in the surrounding
medium by the wind from a young pulsar. In the pulsar wind scenario, the 
rotation of a young pulsar provides stable energy to the nonthermal
relativistic particles in the shocked pulsar wind material outflowing from
the binary companion. As in the microquasar-jet models, the $\gamma$-ray
emission can be produced by inverse Compton scattering of the relativistic
particles from the pulsar wind on stellar photons \citep[][and references
therein]{torres10}.  In the pulsar wind scenario the resolved radio
emission is not due to a relativistic jet akin to those of microquasars,
but arises instead from shocked pulsar wind material outflowing from the
binary \citep{dubus06}.

\begin{figure*}[t]
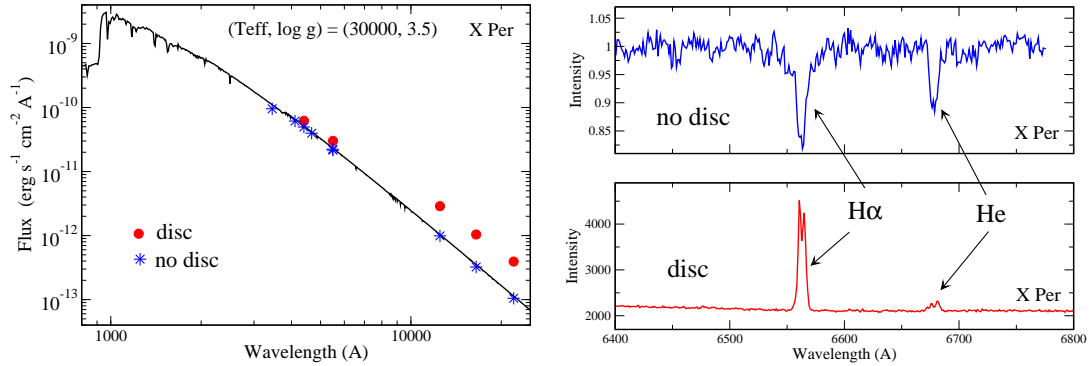

\begin{center}
\begin{tabular}{cc}
\includegraphics[width=0.4\linewidth]{./preig_fig04a.eps} &
\includegraphics[width=0.4\linewidth]{./preig_fig04b.eps} \\
\end{tabular}
\caption{IR excess and H$\alpha$ emission are the two main observational
characteristics of Be stars.}
\label{be}
\end{center}
\end{figure*}

\subsubsection{$\gamma$-Cas like objects}

A growing number of early Be stars discovered in X-ray surveys exhibit
strong X/optical flux correlations and X-ray luminosities intermediate
between those of normal stars and those of most \bex\ in quiescence
\citep{lopes06}. The optical properties are very similar to those of \bex:
{\em i)} the spectral type is always in the range B0-B1 III-Ve, and {\em ii)}
show \ha\ equivalent widths stronger than $-20$ \AA. However, they differ from the
typical \bex\ in their X-ray properties: {\em i)} they show harder X-ray
spectra that can be best fitted by a thin thermal plasma with $T\sim 10^8$
K, rather than a power law as seen typically in HMXBs, {\em ii)}  there is
no evidence for coherent pulsations in any of these systems but strong
variability on time scales as short as 100 s is usually observed, and {\em
iii)} they do not exhibit large X-ray outbursts.

The prototype of this group of sources is $\gamma$ Cas. Two models have
been put forward to explain this type of system: accretion onto a compact
object (most likely a white dwarf) and magnetically heated material between
the photosphere of the B star and the inner part of its disc
\citep{robinson02}.  In the magnetic corona model the hard X-rays result
from high-energy particles that are emitted due to magnetic reconnection,
while the optical variability is due to changes in the density structure of
the inner disc as a consequence of turbulence generated by changes in the
magnetic field. In support of this scenario there is the fact that  X-ray
fluxes  show random variations with orbital phase, thereby contradicting
the binary accretion model, which predicts a substantial modulation.
In favour of the binary model is the similarity of the X-ray spectra with
those of cataclysmic variables, the rapid variability and large numbers of
Be + white dwarf systems predicted by the theory of evolution of massive
binaries.

In the remaining part of this report I shall concentrate on the properties
and variability of \bex\. Only when for the sake of the discussion a
comparison with the behaviour of other type of X-ray binaries may be
illustrative, will these other type of binaries be mentioned.

\section{Optical/IR properties of \bex} 
\label{optir}

The optical and infrared flux of a \bex\ is completely dominated by the Be
star companion. While the X-ray emission from \bex\ provides information on
the physical conditions in the vicinity of the compact object, the optical
and infrared emission reveals the physical state of the mass donor
component. Since the fuel that powers the X-ray emission, namely matter
from a powerful stellar wind or from a equatorially concentrated denser
disc, comes from the massive companion, observations of \bex\ in the
optical/IR are crucial to understand the physical conditions under which
the neutron star is accreting.

The two main observational characteristics of Be stars are the emission
spectral lines, as opposed to the normal absorption photospheric lines and
an excess of IR emission. Both properties, line emission and IR excess,
originate from extended circumstellar envelopes of ionised gas surrounding
the equator of the B star. They result from free-free and free-bound
emission from the disc, i.e. recombination of the optical and UV radiation
from the central star \citep{wool70,gehr74}. The assumption of a common
origin for the line emission and IR excess is strongly supported by the
correlation between the intensities of continuous IR emission, as measured
either by colour indices $(J-M)$, $(J-K)$ or fluxes at a certain wavelength
and the intensity of the H$\alpha$ line, as measured either as equivalent
width or fluxes \citep{dach82,dach88,neto82}. By studying the line
variability one can obtain important constraints on the geometry (size,
shape) and dynamics (velocity and density laws) of the envelope 
\citep{hanu86,hanu95,dach92}.

Figure~\ref{be} illustrates graphically these two properties for the system X
Persei. When an equatorial disc is present the near-IR emission exceeds
that predicted by model atmosphere (shown is a Kuruzc model with $T_{\rm
eff}=30000$ K and $log \, g=3.5$) and emission lines are prominently seen.

\subsection{\ha\ line profiles}
\label{vr}

The H$\alpha$ line is the prime indicator of the circumstellar disc state.
H$\alpha$ emission lines can be morphologically divided in two classes
\citep{humm95,silaj10}: class 1 are symmetric and includes symmetric
double-peak, wine-bottle and shell profiles and class 2, which are
asymmetric and show variability on time scales of years. Each class does
not refer to different groups of sources because individual Be stars can
change from one to the other. These changes are normally slow (of the order
of years to decades in isolated Be stars and months to years in \bex).
Symmetric profiles are believed to be generated in quasi-Keplerian discs
\citep[see e.g.][]{humm94}, while asymmetric profiles are associated with
distorted density distributions \citep{hanu95,humm97}. 

\begin{figure}[t]
\begin{tabular}{c}
\includegraphics[width=0.9\linewidth]{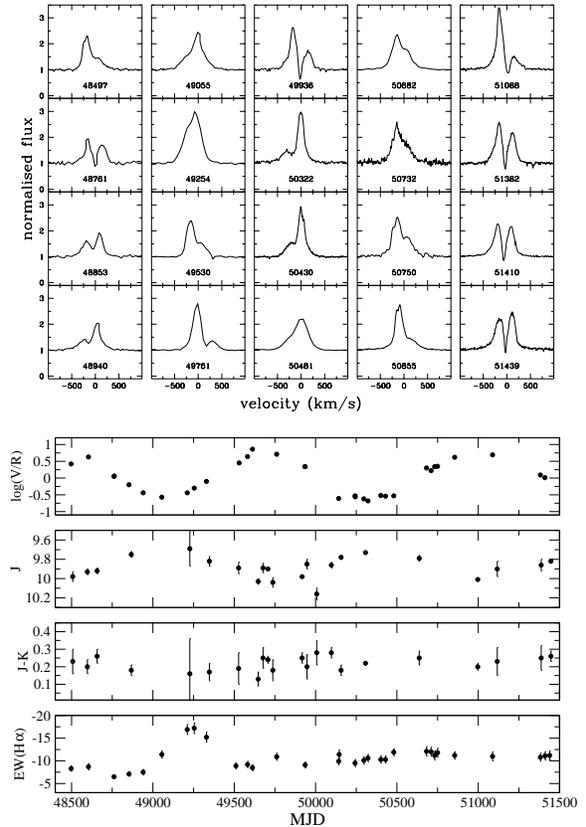} \\
\includegraphics[width=0.9\linewidth]{./preig_fig05b.eps}
\end{tabular}
\caption{{\em Top}: V/R variability observed in the H$\alpha$ line of 
LS I +61 235. From \citet{reig00}. {\em Bottom}: Evolution of the $V/R$ ratio, 
$J$ magnitude, $J-K$ colour and the H$\alpha$ equivalent width in LS I +61 235.
From \citet{reig00}.}
\label{lsi_vr}
\end{figure}

Most \bex\ show asymmetric split H$\alpha$ profiles. The peaks adopt the
names of the relative position of their central wavelengths. In a spectrum
with monotonically increasing wavelength values, the left peak is known as
the "blue" (or violet) peak, while the right peak is named the "red" peak. 
V/R variability refers then to the variation of the relative strength of
the blue to the red peak. Therefore the V/R ratio, defined as the ratio of
violet-side to red-side peak intensities above continuum in units of
continuum intensity represents a measure of the asymmetry of the line. A
more convenient quantity is the logarithmic of this ratio, $\log(V/R)$,
because in this case, positive values of $\log(V/R)$ correspond to a
blue-dominated profile and negative values to a red-dominated line. The top
panel of Fig.~\ref{lsi_vr} displays the V/R variability of the \bex\ LS I +61
235 \citep{reig00}.

V/R variability can be explained in terms of a  non-axisymmetrical
equatorial disc in which a one-armed perturbation  (a zone in the disc with
higher density) propagates \citep{okaz91,okaz96,okaz97,papa92,savo93}.
Double-peak symmetric profiles are expected when the high-density part is
behind or in front of the star, while asymmetric profiles are seen when the
high-density perturbation is on one side of the disc. More precisely,  when
the high-density part of the disc perturbation is located on the side of
the disc where the rotational motion is directed away from us, we see
enhanced red emission (V $<$ R), while when the high-density part is moving
toward the observer, blue-dominated profiles $V>R$ are expected
\citep{telt94}. For systems seen at high-inclination angle, the two
symmetric cases can be readily distinguished since the central depression
between the two symmetric peaks would be more pronounced (reaching or going
beyond the continuum, the so-called shell profile), when the perturbation
is between the observer and the star. The reason is that shell profiles are
thought to be due to a perspective effect, namely, when the line of sight
toward the star probes the equatorial disc and self-absorption is produced
\citep{rivinius06}. If the density perturbation revolves around the star in
the same direction as the material in the disc (prograde precession), then 
the V/R sequence would be according to \citet{telt94}: $V=R$ (perturbation
behind the star) $\longrightarrow$ $V>R$ $\longrightarrow$ $V=R$
(perturbation in front of the star, shell profile) $\longrightarrow$ $V<R$.

One  prediction  of the model is that no  changes  in the slope of the
infrared   continuum  are   expected because  the  V/R variations are not
the result of changes in the radial gradient of the circumstellar  gas. 
This is exactly the behaviour that it is found in LS I +61 235 (bottom
panel of Fig.~\ref{lsi_vr}).  While the  individual  infrared  photometric
bands changed  ($\Delta J \approx \Delta H \approx  \Delta K \sim 0.3$
magnitudes) the infrared colours (e.g. $J-K$) remained unchanged
\citep{reig00}.

V/R variability is also seen in other lines, like HeI 6678\AA.
Since helium lines are generated at smaller disc radii than the hydrogen
lines \citep{humm95, jasc04}, the asymmetry of the HeI line
profiles indicates that the internal changes of the disc are global,
affecting its entire structure.

V/R quasi-periods in \bex\ range from 1-5 yr (Table~\ref{loss}) and are
shorter than those seen in isolated Be stars, which are found in the range
2-11 years with an average of 7 years \citep{okaz97}.

\begin{figure}[t]
\centering
\includegraphics[width=0.9\linewidth]{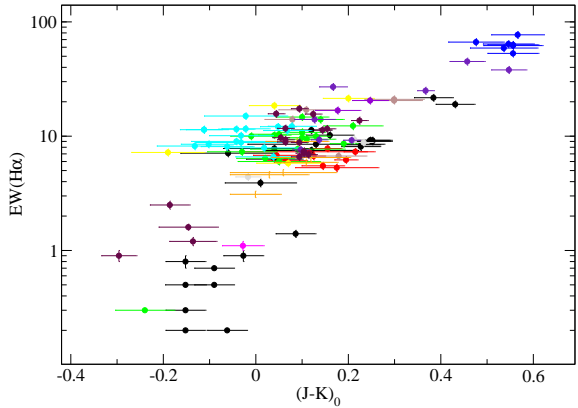}
\caption{Correlation between the equivalent width of the H$\alpha$ line and
the infrared colour $J-K$. Different colours represent different systems. This 
correlation implies a common origin for the
infrared excess and the line emission, namely, the circumstellar disc round
the Be star.}
\label{jkew}
\end{figure}

\begin{table} 
\small
\caption{List of \bex\ used in Fig.~\ref{jkew} with colour excess and amplitude 
of change in $(J-K)_0$ and \ew.}  
\label{jkew_list} 
\begin{tabular}{@{~}l@{~}c@{~}c@{~}c} 
\tableline 
X-ray source	 &$E(B-V)$ &$\Delta (J-K)_0$	  &$\Delta EW(H\alpha)$ \\
		 &(mag)	  &(mag)	  &(\AA)  \\
\tableline
4U 0115+63	 &1.65      &0.8    &20	\\
RX J0146.9+6121	 &0.93      &0.7    &10\\
V 0332+53	 &1.88      &0.4    &6\\
4U 0352+30	 &0.39      &1.1    &18\\
RX J0440.1+4431	 &0.65	    &0.3    &5	\\
1A 0535+262	 &0.75      &1.0    &15	\\
4U 0728-25	 &0.73      &0.1    &2\\
RX J0812.4-3114  &0.65	    &0.2    &5 \\    
GRO J1008-57	 &1.90	    &1.0    &15\\
1A 1118-616	 &0.90	    &1.5    &25\\
4U 1145-619	 &0.29      &2.8    &40\\
EXO 2030+375	 &3.8	    &1.1    &15\\
SAX J2103.5+4545 &1.35	    &0.1    &1	\\
\tableline
\end{tabular} 
\end{table}

\subsection{EW(H$\alpha$) vs infrared colours}
\label{ewir}

If the infrared excess observed in Be stars is due to the same processes as
those responsible for Balmer line emission, namely absorption and
subsequent re-emission of the optical and UV light from the underlying star
in the circumstellar envelope, then a correlation between IR colours and
the strength of the hydrogen lines should be expected. Such correlation has
been reported for isolated Be stars \citep{dach82,neto82,dach88} and \bex\
\citep[][although mixed with isolated Be stars]{coe94,coe05}.
Figure~\ref{jkew} shows the first such diagram made from \bex\ only. It
displays the IR colour index $(J-K)_0$ as a function of the H$\alpha$
equivalent width, \ew, for the sources listed in Table~\ref{jkew_list}. The
infrared colours were corrected for interstellar extinction. Only
contemporaneous data (when the time difference between the IR and optical
observations was less than one month) were used. The second and third
columns in Table~\ref{jkew_list} represent the amplitude of change in the
infrared colour $(J-K)$ and the \ew\ over the observed range, whereas
Fig.~\ref{jkew} plots the actual values.

In Fig.~\ref{jkew} we have included only those sources for which we are
confident that the  colour excess $E(B-V)$ is free of circumstellar
effects. Due to the surrounding envelope, the use of photometric magnitudes
and colours to derive astrophysical parameters, like $E(B-V)$, may be
misleading because they are contaminated by disc emission. The effect of
the disc is to make the photometric indices to appear redder than a
non-emitting B star of the same spectral type. The disc emission makes a
maximum contribution to the optical $(B-V)$ colour of a few tenths of a
magnitude \citep{dach88,howells01}. The values of $E(B-V)$ used to produce
Fig.~\ref{jkew} were obtained from either diffuse interstellar bands,
disc-loss episodes or by disentangling the circumstellar and interstellar
reddening  \citep[see][]{fabr98}. 

Under these conditions, namely, simultaneity of the optical and infrared
data and used of the extinction-corrected $(J-K)$ index, the 
$EW(H\alpha)-(J-K)_0$ becomes a useful tool to estimate the extra reddening
caused by the circumstellar disc. Once a value of the EW(H$\alpha$) is
known, one can look for the corresponding intrinsic $(J-K)$ in
Fig.~\ref{jkew}. By comparing this value with  that expected according to
the spectral type, an estimated of the disc contribution on the index
$(J-K)$ can be obtained. The relatively large scatter, however, may limit
the usefulness of the diagram.

It should be noticed that although the \ha\ and infrared emission
correlate as expected  for a common origin in the disc, the spatial
extension and precise localisation in the disc are different. From
long-baseline interferometric observations in the $K'$ band \citet{gies07}
found that the angular size of the infrared emission is consistently
smaller than that determined for the \ha\ emission. In other words, the
near-IR emission forms closer to the star than does the \ha\ emission. 
Furthermore there is good evidence in several cases
\citep{clar01,grundstrom07} that an increase in disc brightness occurs 
first in the near-IR flux excess and later in \ha\,  as expected for an
outwardly progressing density enhancement. This difference can be explained
if the dependence with disc radius of the \ha\ optical depth is less
steep than that of the infrared optical depth \citep{gies07}.

\begin{figure*}[t]
\begin{center}
\begin{tabular}{cc}
\includegraphics[width=0.3\linewidth]{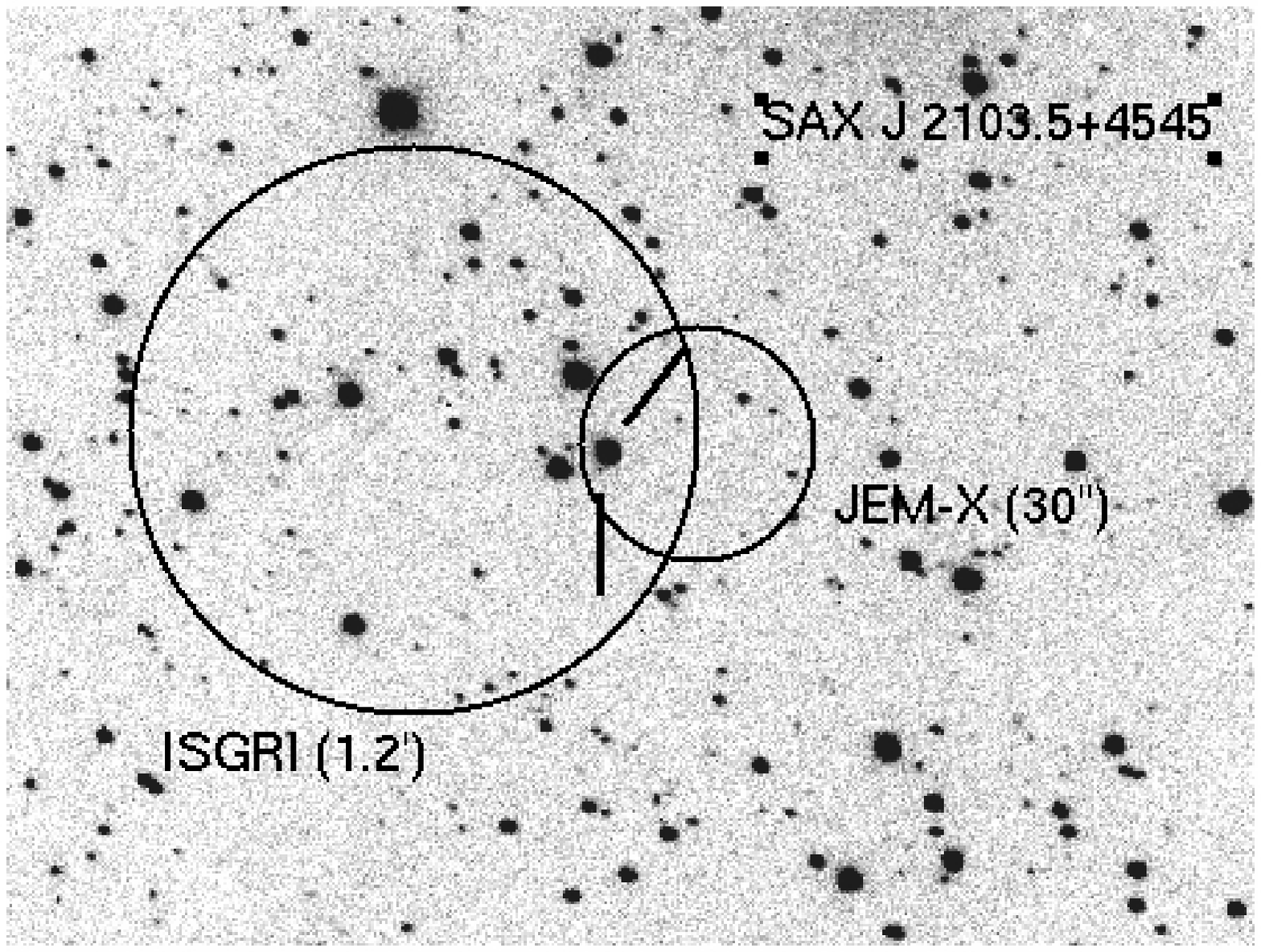} &
\includegraphics[width=0.3\linewidth]{./preig_fig07b.eps} \\
\includegraphics[width=0.3\linewidth]{./preig_fig07c.eps} &
\includegraphics[width=0.3\linewidth]{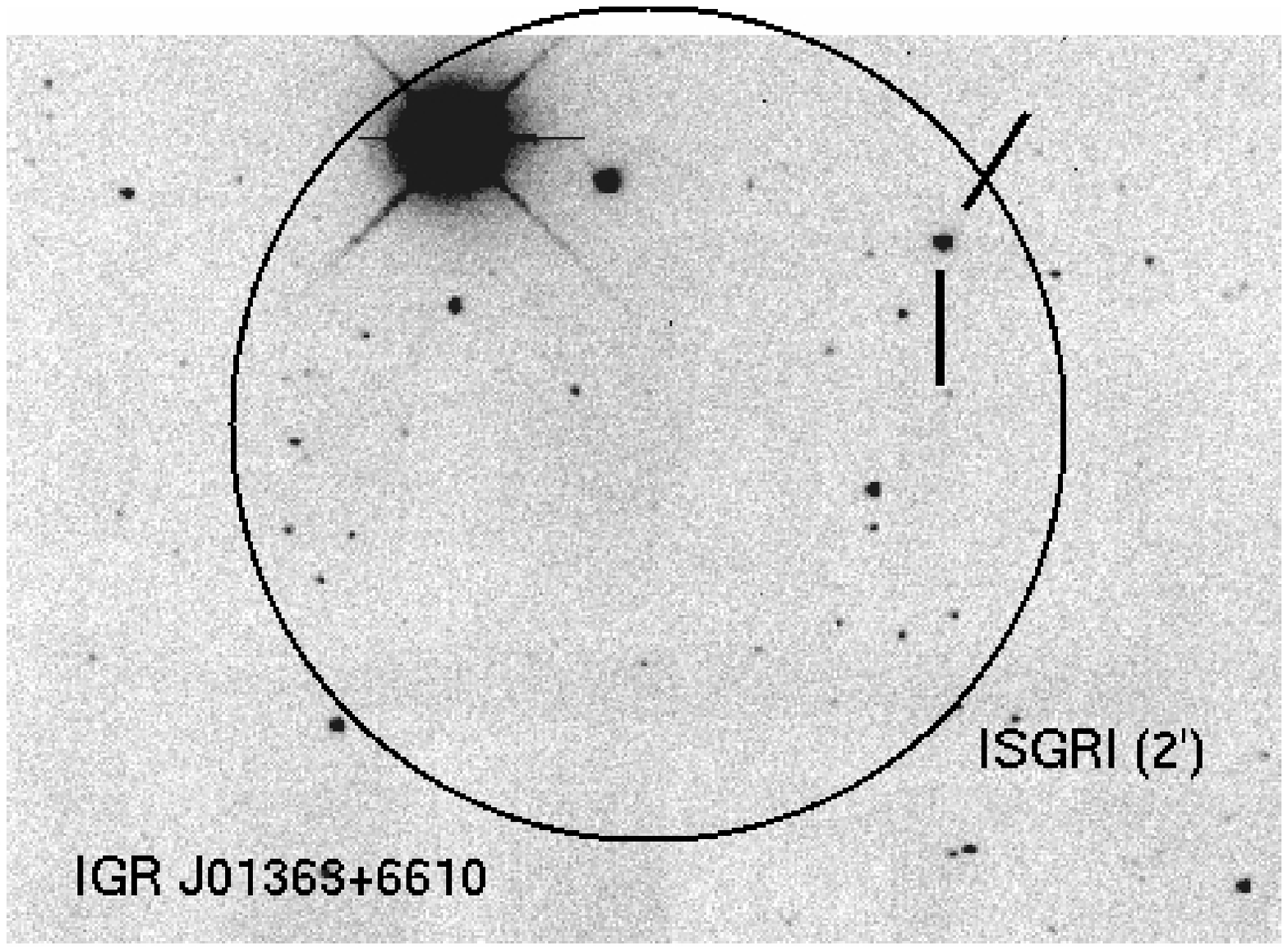} \\
\end{tabular}
\caption[]{Colour-colour diagrams and V-band images of the field around
SAX J2103.5+4545 (top) and IGR J01363+6610 (bottom). The position of the 
confirmed optical counterpart in the colour-colour diagram is marked with a filled circle. 
Good candidates are those that occupy the upper part of the 
colour-colour diagram and lie inside or very close to the X-ray 
uncertainty region. For example, the stars that are close to SAX
J2103.5+4545 in the colour-colour diagram lie far away from the satellite
error circles.}
\label{cdimag}
\end{center}
\end{figure*}
\begin{figure*}[t]
\centering
\begin{tabular}{cc}
\includegraphics[width=0.4\linewidth]{./preig_fig08a.eps} &
\includegraphics[width=0.4\linewidth]{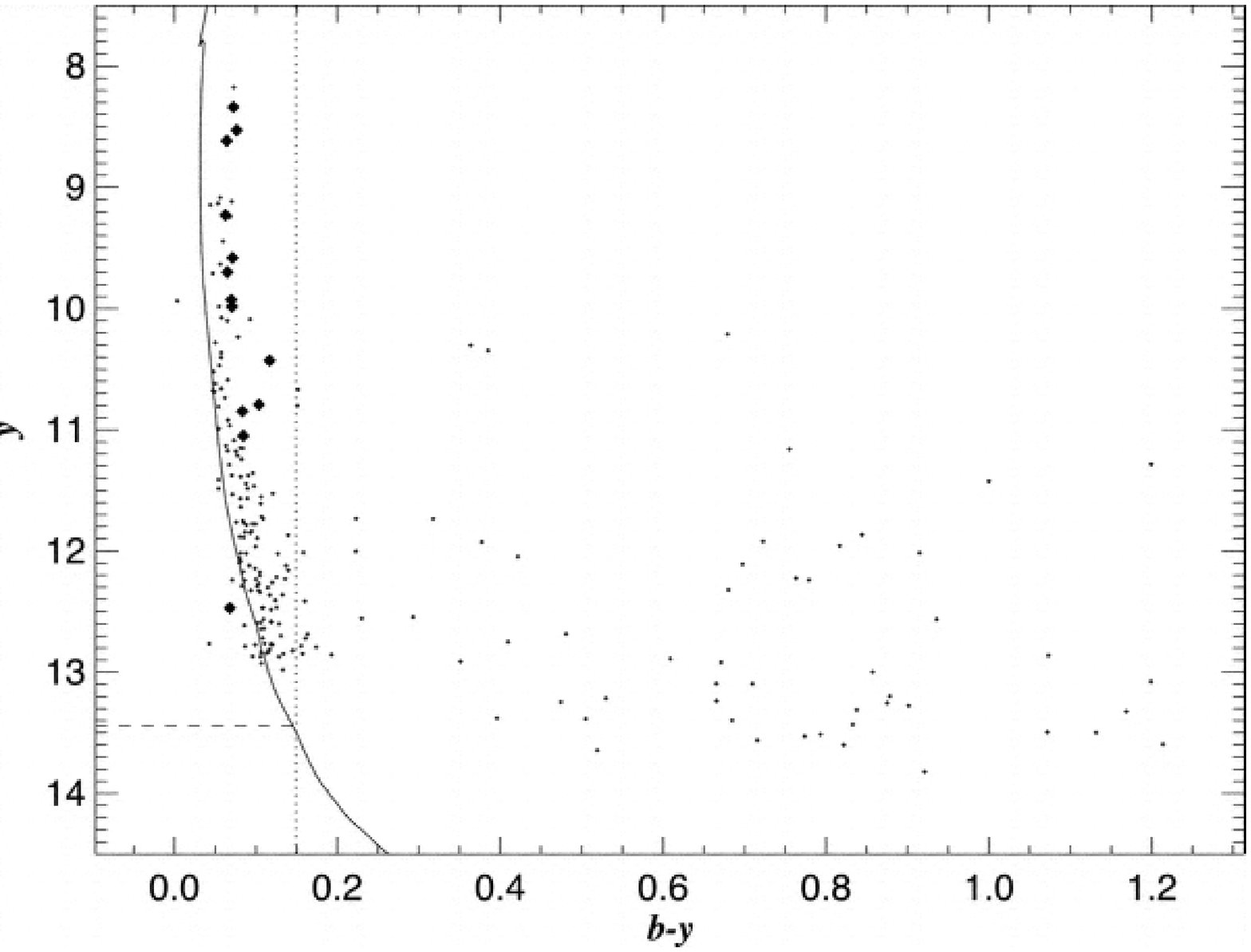}\\
\includegraphics[width=0.4\linewidth]{./preig_fig08b.eps} &
\includegraphics[width=0.4\linewidth]{./preig_fig08c.eps} \\
\end{tabular}
\caption{{\it Top left}: $Q-K$ diagram for three well-known systems. 
\bex\ systems tend to occupy the left-most part of this diagram. {\it Top
left}: ($b-y$,$y$) diagram of the open cluster  NGC 3766. Be stars (diamonds) 
are distinguished from all other stars (dots) in the diagram. From
\citet{mcswain05}. {\it Bottom}: IR photometric diagrams of the field around 
the \bex\ 1A 1118-615 (red cross). \bex\ candidates (open circles) show up 
below the sequence of absorption-line stars (black dots). }
\label{irdiagrams}
\end{figure*}

\subsection{Identification of the optical counterparts}
 
With the improved sensitivities of the currently operational space missions
many new X-ray sources are being discovered. About one third of the the
X/$\gamma$-ray sources in the 4th INTEGRAL catalogue \citep{bird10} are
thought to be X-ray binaries, of which half are believed to contain
early-type companions. An optical identification is necessary to facilitate
a complete study of these systems. Without a known counterpart,
observations are limited to X-ray energies, and hence our understanding of
the structure and dynamics of those systems that remain optically
unidentified is incomplete. 

If the only available information is that provided by an isolated X-ray
detection then potential HMXBs are selected as those exhibiting properties
of a magnetised neutron star, namely, X-ray pulsations  and/or an absorbed
power-law continuum spectrum with an exponential cutoff at 10-30 keV and
cyclotron absorption features. Extra information on the nature of the
source can be obtained if the long-term X-ray variability is known
(information provided by all-sky monitors, like ASM {\it RXTE} or BAT {\it
SWIFT}). The presence of regular and periodic outbursts or unexpected giant
outbursts reaching Eddington luminosity at their peaks may indicate the
presence of a Be star. Likewise, erratic flaring, i.e., X-ray variability
with changes in the X-ray intensity by a factor 3-5 over a few minutes
might indicate the presence of a strong stellar wind from an early-type,
probably evolved, companion.

However, a system with X-ray data only will remain in the category of {\em
suspected} HMXBs until a confirmed optical identification is performed. The
most obvious observational features to look for is H$\alpha$ emission and
near-infrared excess emission. While the detection of these two
observational features does not guarantee that the source is indeed an OBe
type star, it definitely narrows the number of candidates to a handful of
sources.

The size of the uncertainty in the location of the X-ray source, the
so-called error circle, determines the type of observational technique to
use. If the error circle is small (a few arcseconds) then the number of
visible stars in the region is expected to be small and it is possible to
perform narrow-slit spectroscopic observations and look for early-type,
H$\alpha $ emitting line stars. If the error radius is large ($\simmore 1$
arcminute), then it is likely to include a large number of sources and
hence narrow-slit spectroscopy becomes impractical. In this case
colour-colour diagrams can be used to identify good candidates.
Furthermore, uncertainty regions are given to a certain confidence level,
hence it is possible that the true optical companion is located close but
outside the X-ray error circle.

Photometric detection of Be stars can be performed by selecting colours
directly related to the excess emission both in the H$\alpha$ line ("red"
colour) and the continuum ("blue" colour).  Observing through a narrow
filter centred on the H$\alpha$ line and a wider filter also containing
this line, e.g. Johnson R filter \citep{reig05}, Sloan-$r$ \citep{gutt06}
would account for the emission line, while Johnson $B-V$ \citep{reig05} or
Str\"omgren $b-y$ \citep{gutt06} can be used as reddening indicators. 
The use of  Str\"omgren $b$, $y$, and narrow-band  \ha\ photometry through
($b-y$, $y$) colour-magnitude and ($b-y$, $y-H\alpha$) colour-colour diagrams
is particularly suitable to identify Be stars in open clusters
\citep{grebel92,mcswain05}.

Stars with a moderately or large H$\alpha $ excess can be distinguished
from the rest because they deviate from the general trend and occupy the
upper left parts of the diagram. Be star are expected to show bluer
colours, i.e., low $(B-V)$ or $(b-y)$, because they are early-type stars
(although they normally appear redder than non-emitting B stars due to the
circumstellar disc) and also larger (i.e., less negative) $R-{\rm H}\alpha$
colours because they show H$\alpha$ in emission. Once the number of
candidates is reduced to a few systems, narrow-slit spectroscopy becomes
feasible. The final step is to obtain a spectrum in the 4000-5000 \AA, and
perform spectral classification. Figure~\ref{cdimag} shows the V-band
images and the corresponding ($B-V$, $R-H\alpha$) colour-colour diagrams
of two recently identified systems

At other wavelengths, similar photometric diagrams have been used. 
\citet{negu07} defined the reddening-free quantity $Q = (J-H) - 1.70
(H-K_{\rm S})$, and created $Q/K_{\rm S}$ diagrams to separate early type
from late type stars. The majority of stars in Galactic fields concentrate
around Q=0.4-0.5, corresponding to field K and M stars, while early-type
stars typically have $Q\simeq0$. The top left panel in Fig.~\ref{irdiagrams}
shows one such diagram with three well-known \bex.

More recently, \citet{nespoli10} built up infrared colour diagrams
$(Br\gamma-K_{\rm S})-(H-K_{\rm S})$ and $(HeI-K_{\rm S})-(H-K_{\rm S})$ to
identify emission-line objects. Br$\gamma$ is the most prominent feature in
Be stars in the $K$ band, while He I 20580 \AA\ is found in early type Be
stars, up to B2.5 \citep{clark00}. Emission-line objects show up below the
absorption-line stars sequence. Since Be stars present moderate emission
line strengths, when compard to other groups (young stellar objects,
cataclysmic variables, planetary nebulae), Be star candidates are those
below the main sequence but that do not differ more than $\sim 0.3$ mag
(Fig.~\ref{irdiagrams}). The top right diagram in Fig.~\ref{irdiagrams}
also shows an example of the use of narrow-band filters to identify Be
stars in clusters (from \citealt{mcswain05}). 

\section{X-ray properties}
\label{xrayvar}

Tradiationally, \bex\ have been considered as transient X-ray binaries
harbouring a fast rotating neutron star that orbits a Be stype star in a
rather eccentric orbit ($e \simmore 0.3$). Although X-Per was recognized as
a \bex\ since the beginning of the suggestion of the existence of this type
of objects in the early 70's \citep{moff73}, it was the only \bex\ showing
permanent low level X-ray emission. As the number of \bex\ systems began to
grow, the X-ray behaviour increased in complexity and new persistent sources
were discovered \citep{reig99}.

Because all \bex\ are  variable and the detectability of a source depends on
the sensitivity of the detector used,  the  distinction  between transient
and persistent sources undergoing a large intensity increase is sometimes 
ambiguous.  The term {\em  transient}  is normally  used when the 
variability  of the source  exceeds, at least, two orders of magnitude with
respect to the quiescent state, which may correspond to a non-detection
state. 

\begin{figure*}[t]
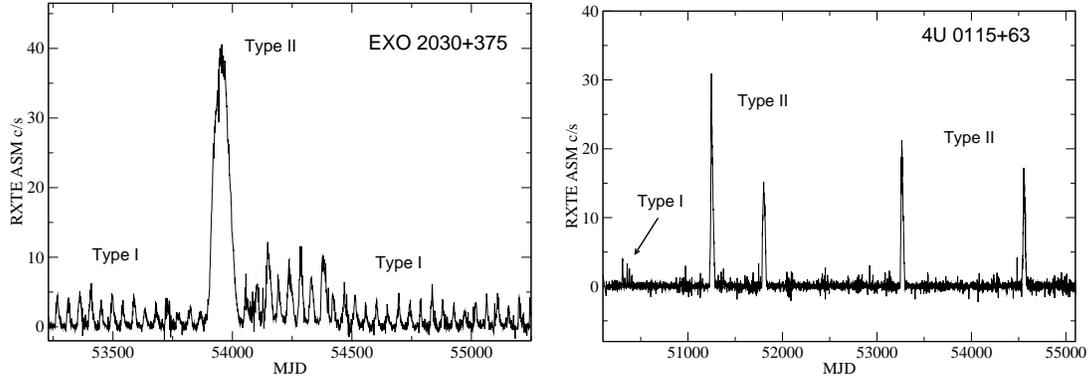

\centering
\begin{tabular}{cc}
\includegraphics[width=0.4\linewidth]{./preig_fig09a.eps} &
\includegraphics[width=0.4\linewidth]{./preig_fig09b.eps} \\
\end{tabular}
\caption{Long-term light curves of EXO 2030+375 (left) and 4U 0115+63
(right). Although the two systems have been seen to display both type I 
and type II outbursts, type II outbursts in EXO 2030+375 are rare 
whereas type I are almost always present. In 4U 0115+63 the opposite occurs.}
\label{outburst}
\end{figure*}

\subsection{Persistent {\it versus} transient X-ray emission}

\citet{reig99} proposed the following characteristics differentiating
between persistent and transient \bex. Persistent sources are characterised
by

\begin{itemize}
\item low X-ray luminosity, $L_{\rm 2-20 keV} \sim 10^{34-35}$ erg
s$^{-1}$.

\item relatively quiet systems showing flat light curves with sporadic and
unpredicted increases in intensity by less than an order of magnitude.

\item slowly rotating pulsars, $P_{\rm spin}\simmore 200$ seconds.

\item absence of, or very weak, iron line at $\sim$6.4 keV, indicative of
only  small amounts of material in the vicinity of the neutron star. 

\end{itemize}

These X-ray properties could be accommodated in systems with  wide ($P_{\rm
orb} \simmore 200 $ days) and low-eccentric orbits ($e\simless 0.2$).  A
thermal excess of blackbody type, with high temperature ($kT > 1$ keV) and
small emission area ($R < 0.5$ km), has recently been suggested as another
common feature of this type of sources  \citep[][and references
therein]{lapa09}. Members of this group are: X Per,  RX J0146.9+6121/LS I
+61 235, RX J0440.9+4431/BSD 24-491 and RX J1037.5-564/LS 1698.

\begin{figure*}[t]
\centering
\includegraphics[width=0.6\linewidth,height=0.5\linewidth]{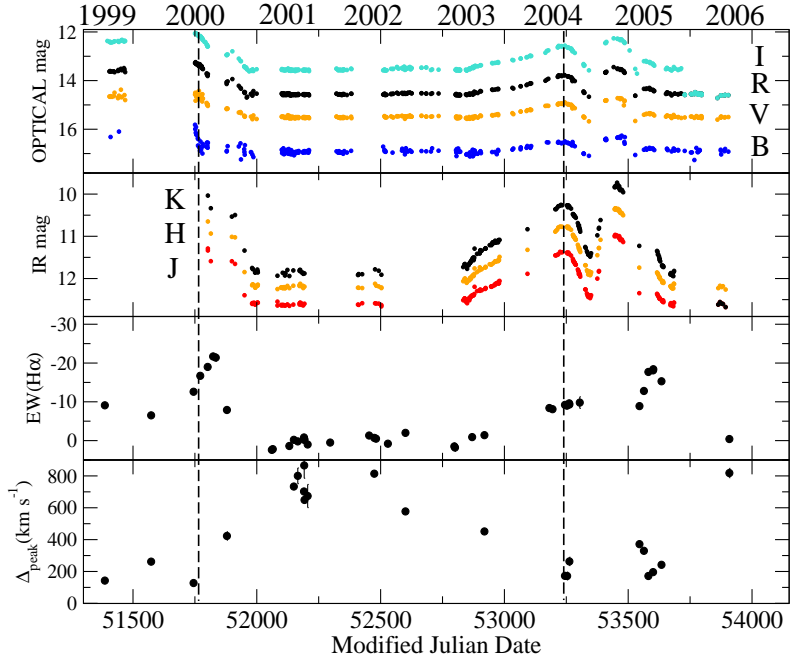} 
\caption[]{Long-term evolution of the optical/IR magnitudes and the 
\ha\ equivalent width and peak separation of double-peak
profiles of 4U 0115+63 . Dotted lines mark the occurrence 
of X-ray outbursts. Adapted from \citet{reig07}}
\label{typeII}
\end{figure*}

In contrast, the X-ray behaviour of transient \bex\ is characterised by two
type of outbursting activity:

\begin{itemize}

\item Type I outbursts. These are regular and periodic (or quasiperiodic)
outbursts, normally peaking at or close to periastron passage of the
neutron star. They are short-lived, i.e., tend to cover a relatively small
fraction of the orbital period (typically 0.2-0.3 $P_{\rm orb}$). The X-ray
flux increases by about one order of magnitude with respect to the
pre-outburst state, reaching peak luminosities $L_x \leq 10^{37}$ erg
s$^{-1}$.

\item Type II outbursts represent major increases of the X-ray flux,
$10^{3}-10^{4}$ times that at quiescence. They reach the Eddington
luminosity for a neutron star and become the brightest objects of the
X-ray sky. They do not show any preferred orbital phase and last for a
large fraction of an orbital period or even for several orbital periods.
The formation of an accretion disc during Type II outbursts 
\citep{kris83,motc91,haya04,wils08} may occur. The discovery of
quasi-periodic oscillations in some systems 
\citep[][and reference therein]{James10} would
support this scenario. The presence of an accretion disc also helps explain
the large and steady spin-up rates seen during the giant outbursts, which
are difficult to account for by means of direct accretion.

\end{itemize}

Figure \ref{outburst} shows these two types of X-ray variability. The
long-term X-ray emission of EXO 2030+375  appears clearly modulated by the
orbital period of 46 days just before and after the type II outburst at MJD
53950. 4U 0115+63 has exhibited four type II outburst in the time interval 1996-2009.

Type II outbursts are major events. Since the fuel that powers these
outbursts comes from the material in the circumstellar disc, one would
expect major changes in the structure of the disc. The disruption should be
observable in the parameters that best sample the physical condition in the
disc, namely the H$\alpha$ line parameters (strength and shape) and the IR
colours and indices. Figure~\ref{typeII} displays the long-term evolution
of the optical/IR and  \ha\ equivalent width of 4U 0115+63. Vertical dotted
lines mark the occurrence  of type II X-ray outbursts. As it can be seen,
the large amplitude changes in the optical and IR photometric and
spectroscopic parameters after the X-ray outbursts indicate major
disruption in the physical conditions of the circumstellar. In fact, the
type II outbursts led to the complete disappearance of the disc. Evidence
for the loss of the disc stems from the H$\alpha$ equivalent width, which
changed sign from negative, indicating that the spectral line was in
emission before the outburst to positive, that is absorption, after the
outburst.

\begin{figure*}[t]
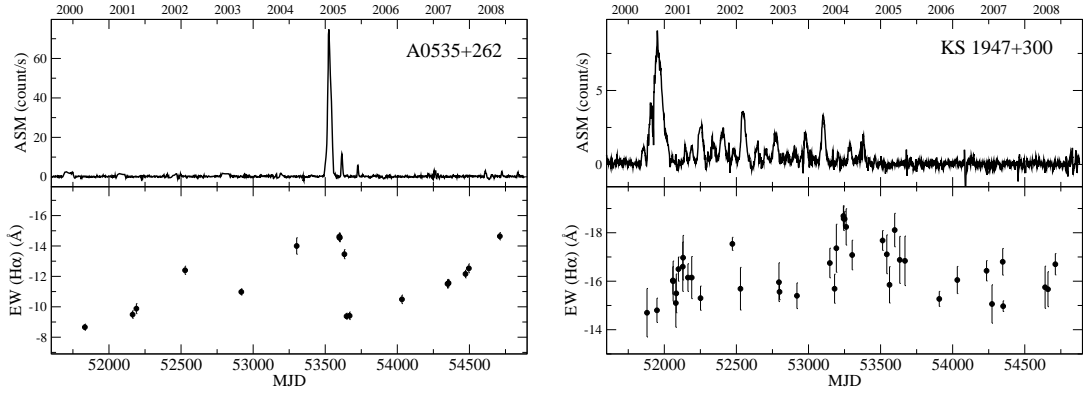

\centering
\begin{tabular}{cc}
\includegraphics[width=0.4\linewidth]{./preig_fig11a.eps}&
\includegraphics[width=0.4\linewidth]{./preig_fig11b.eps} \\
\end{tabular}
\caption[]{Long-term evolution of the X-ray intensity (1.3--12.1 keV) and 
the \ha\ equivalent of 1A 0535+262 (left) and KS 1947+300 (right).}
\label{typeIIb}
\end{figure*}

However, type II outbursts are not always followed by a disc-loss phase.
Figure ~\ref{typeIIb} shows the case of 1A 0535+262 and KS 1947+300. In the
case of 1A 0535+262, the H$\alpha$ equivalent width decreased significantly
after the X-ray outburst, indicating a major disruption in the structure of
the disc, but the H$\alpha$ equivalent width remained negative. The case of
KS 1947+300 is hard to explain. Apparently, the X-ray outbursts produced no
effect on the disc. The H$\alpha$ equivalent width did not change
significantly after the outburst.

\subsection{Periodic variability: X-ray pulsations}

Observations of X-ray binaries show considerable variability on a wide
range of time scales in all wavelengths. In LMXB, in the so-called
millisecond pulsars, the X-ray periodic variability goes down to less than
a millisecond. In \bex\ the shortest timescales are of the order of few
seconds\footnote{There are at least two systems, A0535-66 and SAX
J0635+0533, with B-type companions that exhibit millisecond pulsations.
However, they are thought to be rotation-powered pulsars.}, and correspond to
the X-ray pulsations.

The detection of coherent pulsations from an accreting X-ray source
provides one of the strongest evidence that the compact object is a neutron
star. The pulses of high-energy radiation are due to a misalignment of the
neutron star's rotation axis and its magnetic axis. Gas is accreted from
the stellar companion and is channeled by the neutron star's magnetic field
onto the magnetic poles producing two or more localized X-ray hot spots.
These hot spots move into and out of view as the neutron star spins, giving
rise to regular X-ray pulses.

Almost all \bex\ show X-ray pulsations. The pulsation period of
an X-ray pulsar with a Be companion can be as short as a few seconds (4U
0115+63) and as long as few tens of minutes (LS I +61 235). The longest
periods in main-sequence HMXBs are those of 4U 2206+54 with 1.54 hr
\citep{reig09} and IGR J01583+6713 with 5.47 hr \citep{wang09}. However,
these two sources are somehow peculiar and cannot be considered as
canonical \bex. Although 4U 2206+54 exhibits a double-peaked emission
H$\alpha$ profile and the optical companion is an O9.5 main-sequence star,
its optical spectrum is complex and far from that of a "typical" Be star
\citep{blay06}. The 5.47-hr period of IGR J01583+6713 needs confirmation as
a shorter (469 s) periodicity has also been suggested \citep{kaur08}. Among
HMXBs with a supergiant companion, 2S 0114+65 shows the longest spin period
with 2.78 hr \citep{finl94}. The only \bex\ for which no pulsations have been
detected is LS I +61 303/2E 0236.6+6101. The nature of this system is under
debate. The lack of pulsations can be explained if it contains a black
hole, that is, if LS I +61 303 is a microquasar. Radial velocity studies do
not completely rule out this possibility \citep{casa05a}. On the other
hand, radio observations favour the presence of a neutron star. However, in
this case LS I +61 303 would contain a rotation-powered pulsar with nebula and
not an accretion-powered pulsar \citep[][see also Sect.~\ref{grb}]{dhaw06}.

\begin{figure}[h]
\begin{tabular}{cc}
\includegraphics[width=0.4\linewidth]{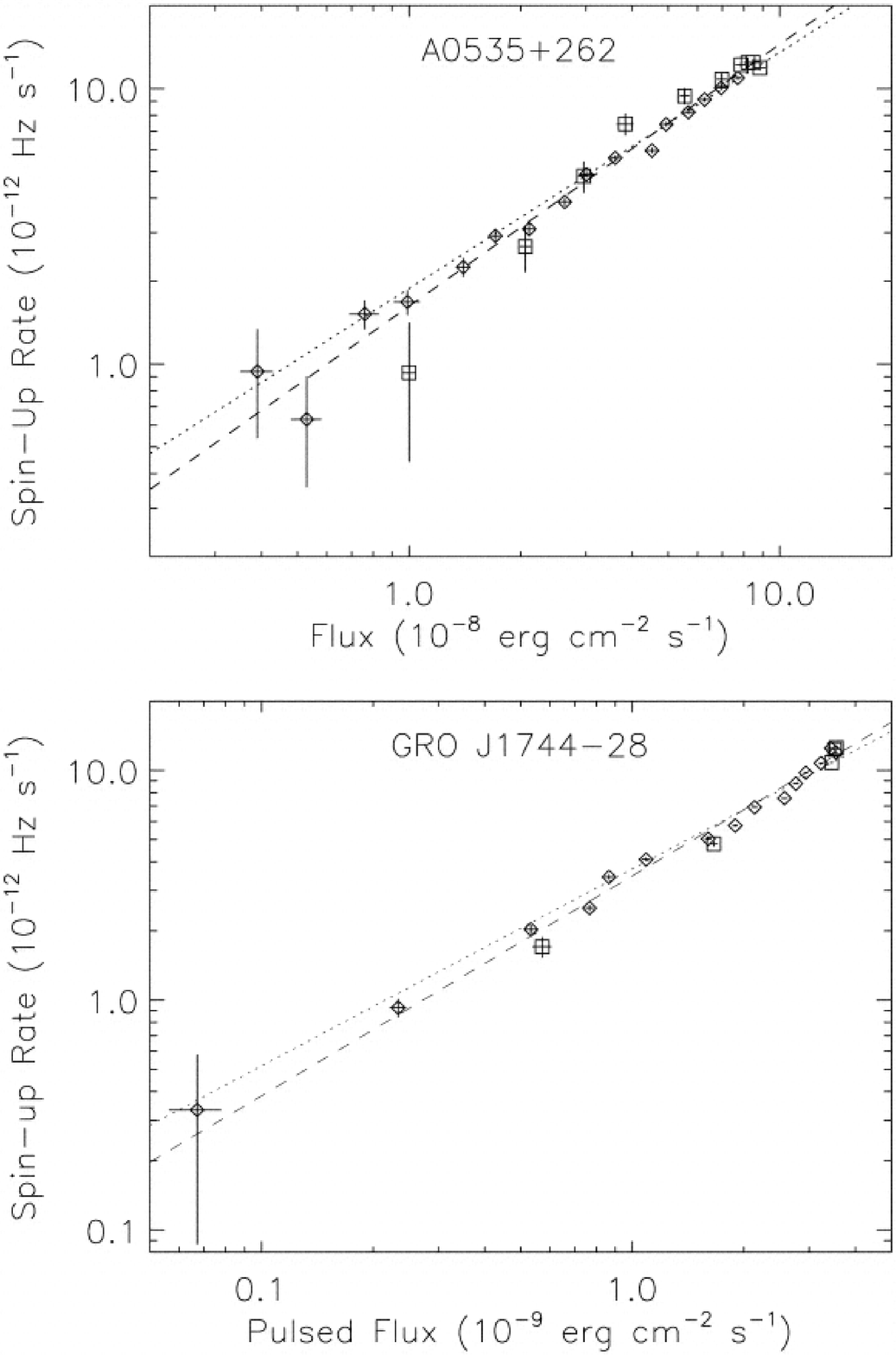}&
\includegraphics[width=0.4\linewidth]{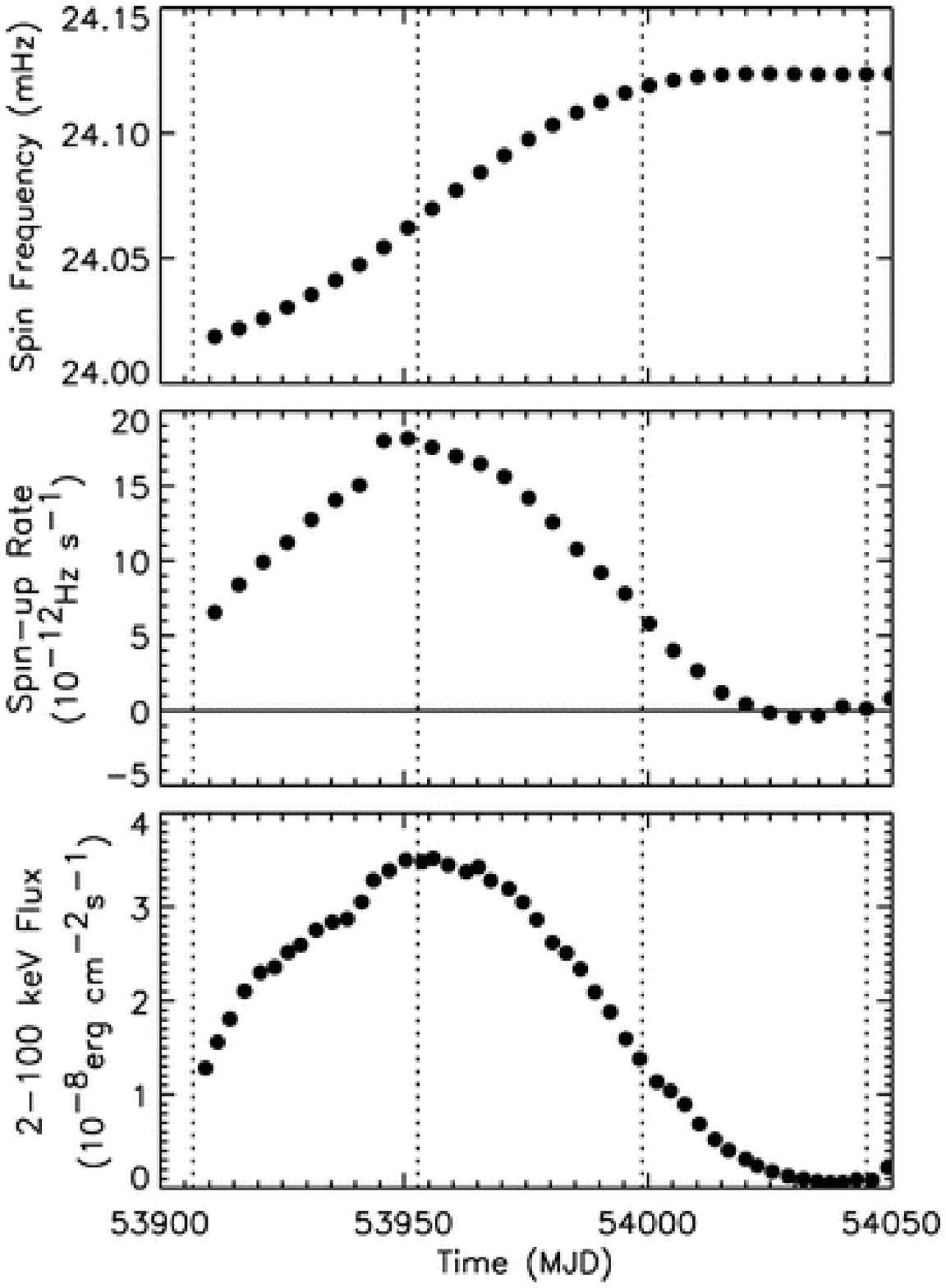} \\
\end{tabular}
\caption[]{Observed relationships between flux and pulsar spin-up rate
{\it Left}: A 0535+262 during the 1994 giant and GRO J1744+28 during the 1995-1996 
outburst \citep{bild97}. {\it Right}: EXO 2030+375 during the 2006 giant
outburst \citep{wils08}}
\label{spinup}
\end{figure}

During major outbursts, transient accretion-powered pulsars show spin-up
episodes, with the spin-up rate increasing with accretion luminosity
(Fig.~\ref{spinup}). This is consistent with the simple model of
disc magnetosphere interaction \citep{ghos79}, in which, as the
accretion rate increases, the rate of angular momentum transfer from the
accretion flow to the neutron star increases. At the limit of very low mass
accretion rate, the neutron stars are expected to spin down, because the
magnetic field lines that couple to the outer, slower accretion flow remove
spin angular momentum from the neutron star.

Spin periods in the range 1-1000 s can be explained according to the
current understanding of the neutron star's spin evolution in a close
binary system \citep{davi79,davi81,zhan04b,dai06}. Longer periods are,
however, difficult to accommodate in the standard theory. The spin
evolution of a neutron star in a binary system is divided into three basic
phases or states. Each state is characterised by a different energy
release mechanism and corresponds to a different evolutionary stage of the
neutron star. These phases are known as the pulsar (or ejector) phase, the
propeller phase and the accretion phase.  To explain the existence of
long-period pulsars ($P_{\rm spin}\simmore 1000$ s), a subphase of the
propeller state needs to be invoked. This subphase is called the subsonic
propeller phase \citep{davi81,ikhs01a,ikhs01b,ikhs07}.

After a supernova explosion a neutron star is formed as a rapidly rotating
radio pulsar. The spin period at birth is of the order of a fraction of a
second and its magnetic field of the order of $10^{13}$ G. In this phase,
the neutron star radiates at the expense of its rotational energy.

The condition for the transition of a slowing-down neutron star from the
ejector phase into the supersonic propeller phase occurs when the pressure
of the relativistic wind ejected by the neutron star (magnetic pressure) is
balanced by the pressure of the plasma surrounding the neutron star (ram
pressure) at the gravitational capture or accretion radius $r_{\rm
acc}=2GM_x/v_w^2$, where $M_x$ is the neutron star mass and $v_w$ is the
stellar wind velocity. The plasma will tend to accrete onto the neutron star
under the action of gravitation. The electromagnetic field, however, will
obstruct this process, and the accreting matter will come to a stop at the
magnetospheric boundary. The accreted material cannot penetrate any further
because the drag exerted by the magnetic field is super-Keplerian. The
in-falling matter is accelerated outward, taking away the angular momentum
of the neutron star. Thus the spin-down of the neutron star continues until
the centrifugal and gravity forces acting on a particle that corotates with
the neutron star balance. This occurs at the so-called equilibrium period.

If the neutron star spin period is larger than the equilibrium period, then the
centrifugal barrier ceases to be effective and plasma that penetrates into
the magnetosphere is able to reach the neutron star surface, moving along
the magnetic field lines in the direction of the magnetic poles.

Since the material accreted carries substantial angular momentum, the
neutron star will experience strong spin-up. Therefore, in principle, the
equilibrium period given by 

\[P_{\rm eq} \approx  23 \mu_{30}^{6/7}\, \dot M_{15}^{-3/7}
\left(\frac{M_x}{\msun}\right)^{-5/7} \; \; \; {\rm s} \]

\noindent represents the maximum period for a given mass accretion rate. 
$\mu_{30}$ and $\dot M_{15}$ are the magnetic moment and mass accretion
rate in units of $10^{30}$ G cm$^{3}$ and $10^{15}$ g s$^{-1}$,
respectively. 

$P_{\rm eq} \sim 10^3$ s can be achieved if the magnetic field strength in
units of $10^{12}$ G, $B_{12}> 100$ or if $\dot{M}_{15}< 10^{-3}$. None of
these conditions is supported by the observations. The condition of
supercritical magnetic fields would imply that the neutron star is a
magnetar. However, the detection of cyclotron lines in the range 10-100 keV
indicates that the magnetic field strength is of the order a few times
$10^{12}$ G ($B_{12}\sim 1-6$). The low mass accretion rate condition would
imply an X-ray luminosity two to three orders of magnitude lower than
observed.

In order to explain spin periods longer than $\sim10^3$ s, another phase in
the spin-period evolution has to be considered. This phase is called
subsonic propeller phase and occurs prior to the accretion-powered
phase \citep{ikhs01a,ikhs01b,ikhs07}.

We indicated above that if $P_{\rm spin} > P_{\rm eq}$ then matter can be
accreted onto the neutron star surface. However, this is actually not
always true. \citet{davi81} showed that the star's magnetosphere during the
propeller spin-down epoch is surrounded by a hot, spherical quasi-static
envelope. This envelope extends from the magnetospheric radius
$r_m = \left(\mu^2/\dot M\sqrt{2GM_x}\right)^{2/7}$
up to the accretion radius $r_{\rm acc}=2GM_x/v_{w}^2$.
The interaction between the magnetosphere and the envelope leads to the
deceleration of the rotation rate of the neutron star: the rotational
energy loss by the star is convected up through the envelope by the
turbulent motions and lost through its outer boundary. The neutron star
remains in the propeller state as long as the energy input to the envelope
due to the propeller action by the star dominates the radiative losses from
the envelope plasma. 

Steady accretion under the condition $P_{\rm spin} > P_{\rm eq}$ can be
realized only if the cooling of the envelope plasma (due to radiation and
convective motions) dominates the energy input. The subsonic propeller
phase corresponds to the phase when this cooling occurs.

The condition for accretion then translates to  $P_{\rm spin} > P_{\rm
br}$, where $P_{\rm br}$ is the so-called break period, which is
given by \citep{ikhs07}

\begin{eqnarray*}
P_{\rm br}&\approx& 7\times 10^{-4} \mu^{16/21} M_x^{-4/21} \dot M^{-5/7} \\
P_{\rm br}&\approx& 442 \; \mu_{30}^{16/21} \dot M_{15}^{-5/7}
\left(\frac{M_x}{\msun}\right)^{-4/21} \; \; \; {\rm s}
\label{break}
\end{eqnarray*}

\noindent While $P_{\rm spin} < P_{\rm br}$ the temperature of the envelope
remains higher than the free-fall temperature ($T_{\rm ff}(r_m)=GM_x
m_p/kr_m$, where $m_p$ is the proton mass and $k$ the Boltzmann constant)
and no accretion is possible.

\begin{figure*}[t]
\centering
\includegraphics[width=0.8\linewidth]{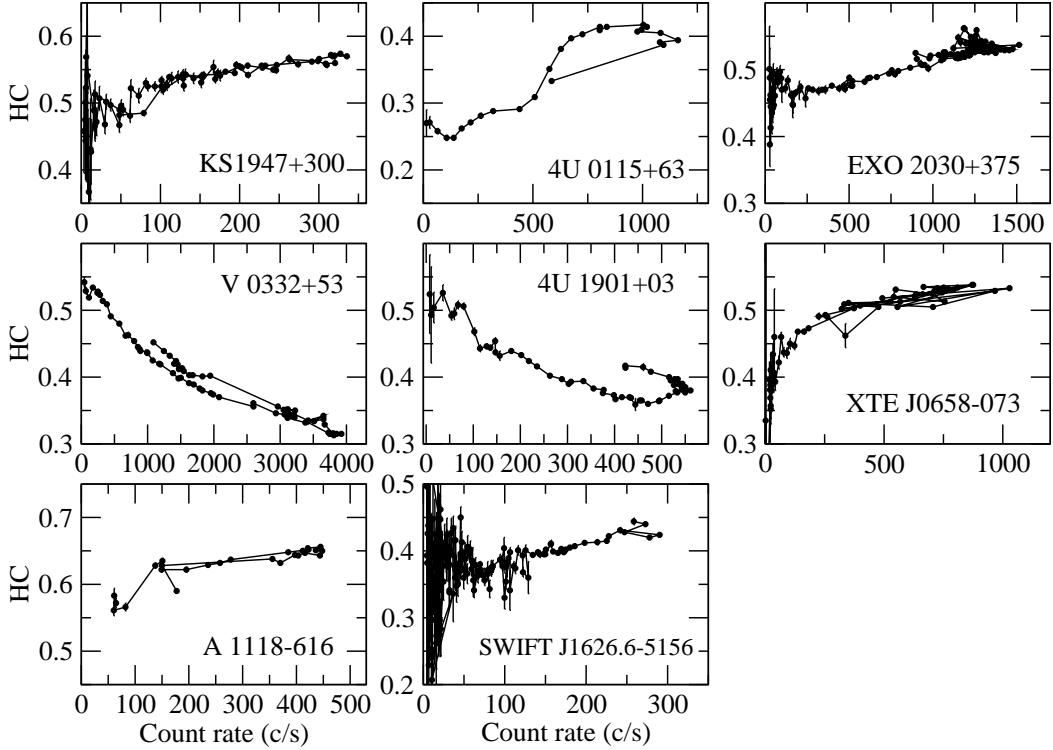}
\caption[]{Hardness-intensity diagrams traced by various \bex\ during giant
(type II) outbursts observed by {\it RXTE}. The {\em hard colour}, HC, is 
the ratio between the count rate in the energy range 15-30 keV over 10-15 keV. }
\label{hid}
\end{figure*}

\subsection{Aperiodic variability}

Aperiodic variability refers to the variability caused by processes that
do not repeat themselves in a periodic way \citep{klis95,klis06}. Pulsations, eclipses,
outbursts are excluded from the analysis of aperiodic variability studies.
In contrast, flickering, irregular flaring, fluctuations are the subject of
study. The main tool used for studying the aperiodic variability is
the Fourier  power spectrum, or a Power Spectral Density (PSD), of the
count-rate time series. A number of variability components, or
power-spectral components, together  make up the power spectrum. An
aperiodic component by definition covers several frequency resolution
elements. Broad features are called noise and narrow features are called
quasi-periodic oscillations (QPOs). The so-called $Q$-factor determines
whether a noise component deserves the name QPO. If the $Q$-factor,
defined as the ratio between the characteristic frequency of the component
over its width, $Q=\nu/FWHM$, is larger than 2, then the noise component is
considered to be narrow enough to be called a QPO.

Aperiodic variability studies are normally carried out in correlation with
the position of the source in the colour-colour diagram and/or
hardness-intensity diagram.  An X-ray color is a 'hardness' ratio between
the photon counts in two broad bands.  A colour-colour diagram (CD) is the
plot of one color vs. another calculated in different energy bands. The
higher energy range color is called {\em hard colour}, while the lower
energy range is the {\em soft colour}.  A hardness-intensity diagram (HID)
is instead the plot of a colour vs. the count-rate in some broad spectral
range. Figure~\ref{hid} shows the HID of eight \bex\ during type II
outbursts. The CD/HID are very useful tools to investigate the spectral
variations of an X-ray source because they are model- and
instrument-independent and reflect the intrinsic properties of the system.

Together with the functional form of the variability components ("noise"),
CD and HID have revolutionised our understanding of X-ray
binaries, unveiling a whole new phenomenology on the spectral and timing
properties of X-ray binaries by introducing the notion of source states.  A
state is defined by the appearance of a spectral (i.e. power-law,
blackbody) or variability component (i.e. Lorentzian) associated with a
particular and well-defined position of the source in the CD/HID. A
transition between states takes place when the relative strength of the
spectral or variability components varies and the source motion in the
CD/HID changes direction \citep{klis06}.

In LMXBs and BHBs the aperiodic variability is thought to originate in the
irregular nature of the inner accretion flow, as an accretion disc is
assumed to exist in this type of systems. In HMXBs the details of how the
magnetic field affects the accretion flow are not fully understood.

While there are numerous references in the literature on the application of
CD/HID analysis on LMXBs and BHBs, very little work of this type has been
done on HMXBs \citep{bell90}. Recently, \citet{reig08} performed the first
systematic analysis of the spectral and rapid aperiodic variability of
Be/X-ray binaries during type II outbursts. He found similarities and
differences with respect to the behaviour of LMXBs and BHBs.

\begin{figure*}[t]
   \centering
   \includegraphics[width=0.7\linewidth]{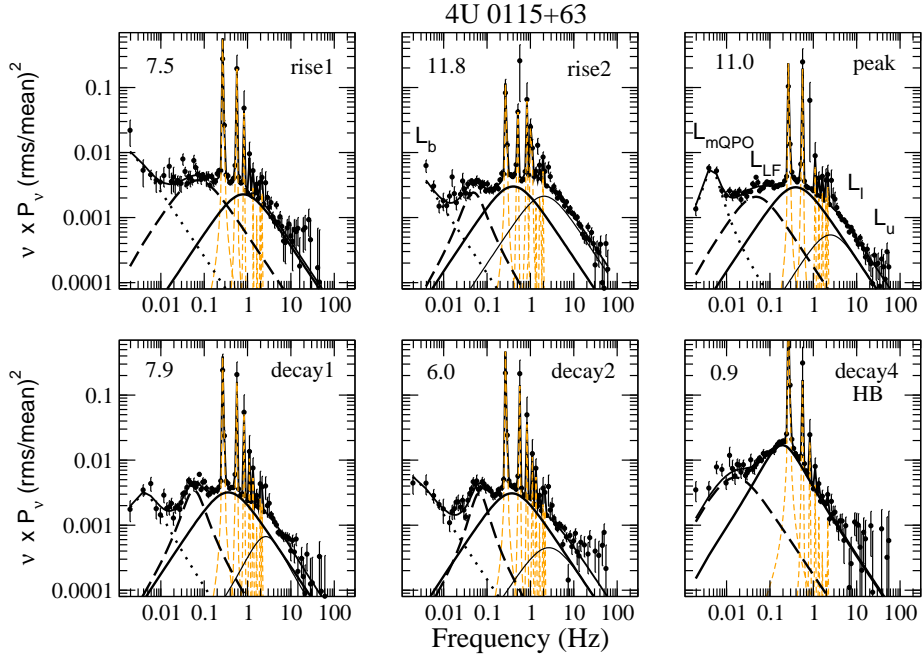} \\
      \caption{Power spectra of 4U 0115+63 at various stages of the 2004 giant
      outburst. The number at the top
      left of each panel is the X-ray luminosity in units of $10^{37}$ erg
      s$^{-1}$. From \citet{reig08}.} 
      \label{psd}
   \end{figure*}

The similarities include: {\em i)} existence of spectral branches in the
CD/HID. At high and intermediate flux the sources move
along a diagonal branch (DB), while at very low flux the soft colour
decreases while the hard colour remains constant defining a horizontal
branch (HB); {\em ii)} the source does not jump through the diagram but
moves smoothly, i.e. without jumps, following the pattern; {\em iii)} the
broad noise components in the HB are more variable (in terms of fractional
rms) than in the DB; {\em iv)} the power spectra of HMXBs can be fitted
acceptably with a sum of a small number (2--4) of Lorentzian functions
(Fig.~\ref{psd}). All sources present peaked noise in certain spectral
states. 

While the differences are: {\em i)} presence of hysteresis patterns in the
CD/HID that might be a manifestation of the magnetic field (through
cyclotron resonant scattering features); {\em ii)} slower motion along the
spectral branches, hours to days in LMXBs, days to weeks in HMXBs; {\em
iii)} the characteristic time scales implied by the noise components are
about one order of magnitude longer in HMXBs. There are about 14 HMXBs
pulsars that show QPOs in their power spectra \citep{james10}. The
frequencies of these QPOs range from 1--200 mHz. In
contrast, in LMXBs and BHBs QPO range 1-1000 Hz; {\em iv)} the so-called
$L_S$ noise, whose central frequency coincides with the frequency of the
fundamental peak of the pulse period and suggests a strong coupling between
the periodic and aperiodic noise components. This coupling implies that the
instabilities in the accretion flow that give rise to the aperiodic
variability must travel all the way down to the neutron star surface
\citep{lazz97,burd97}, and {\em v)} no apparent correlation between the
power spectral parameters (characteristic frequencies, $rms$) and mass
accretion rate.

\begin{table*}
\begin{center}
\caption{Rotational velocity of Be/X-ray binaries}
\label{rotvel}
\begin{tabular}{lllccc}
\tableline
X-ray		&Optical	&Spectral      &Inclination    &$v \sin i$     &Reference \\
source		&counterpart	&type	       &angle ($^{\circ}$)&(km s$^{-1}$)& \\
\tableline
4U 0115+634	&V635 Cas	&B0.2V	   &43     &300$\pm$50     &1 \\
RX J0146.9+6121	&LS I +61 235	&B1III-V   &--	   &200$\pm$30     &2 \\
V 0332+53	&BQ Cam		&O8-9V	   &$<$10  &$<$150	   &3 \\
X-Per		&HD 24534	&O9.5III   &23-30  &215$\pm$10     &4,5 \\
RX J0440.9+4431	&LS V +44 17	&B1III-V   &	   &235$\pm$15     &6 \\
1A 0535+262	&HD 245770 	&O9.7III   &28-35  &225$\pm$10     &7,8 \\
RX J0812.4--3114&LS 992		&B0.5III   &--	   &240$\pm$20     &9 \\
1A 1118--615	&Hen 3-640	&O9.5IV	   &--	   &300 	   &10 \\
4U 1145--619	&V801 Cen 	&B0.2III   &$<$45  &250$\pm$30, 290&11,12 \\
4U 1258--61	&V850 Cen 	&B2V	   &90 sh  &--		   &13 \\
SAX J2103.5+4545&--		&B0V	   &--	   &240$\pm$20     &14 \\
\tableline
\end{tabular}
\tablecomments{[1] \citet{negu01}, [2] \citet{reig97}, [3] \citet{negu99}
[4] \citet{lyub97}, [5] \citet{delg01}, [6] \citet{reig05}
[7] \citet{haig04}, [8] \citet{grun07}, [9] \citet{reig01}
[10] \citet{jano81}, [11] \citet{jano82}, [12] \citet{webs74}
[13] \citet{park80}, [14] \citet{reig04}}
\end{center}
\end{table*}

\section{Circumstellar disc}

\subsection{Disc formation}
\label{discform}

Be stars are fast rotators. They have, on average, larger observed
rotational velocities than B stars as a group \citep{slet82}. The
determination of the rotational velocity is believed to be a crucial
parameter in the formation of the circumstellar disc. A rotational velocity
close to the break-up or critical velocity (i.e. the velocity at which
centrifugal forces balance  Newtonian gravity) reduces the effective
equatorial gravity to the extent that weak processes such as, gas pressure
and/or non-radial pulsations may trigger the ejection of photospheric
matter with sufficient energy and angular momentum to make it spin up into
a Keplerian disc.

At present, there is no consensus on how close to critical velocity Be
stars rotate, nor on how Be stars reached such high spinning rates.
Observations suggest that a large fraction of Be stars rotate at 70--80\%
of the critical value \citep{slet82,porter96,yudi01}.  However,
\citet{town04} claimed that if gravity darkening is taken into account then
most or all Be stars may be rotating very nearly at critical velocity.
Whether the rotational velocities of Be stars reflect the initial
distribution of angular momentum or they have been spun up as a result of
binary evolution is still a matter of debate.  There is growing evidence
that supports the idea that some Be stars were spun up by mass transfer in
interacting binaries \citep{gies00}. The former mass donors  are now found
as neutron stars (i.e. \bex) in some cases and as a small hot subdwarfs
stripped of its outer envelope (Phi Per, 59 Cyg, FY CMa) in other cases
\citep{gies98,rivinius04,maintz05,peters08}.   

The projected rotational velocities are determined by measuring the width
of certain spectral lines \citep[see e.g.][]{steele99}, as it is generally
assumed that the width of the spectral lines and the rotational velocity
are linearly related. This is true even with the inclusion of limb and
intrinsic darkening, but if gravitational darkening is taken into account,
then for stars rotating above $\sim$80\% of critical speeds the linear
relationship is no longer applicable. Any increase in the rotational
velocity is accompanied by almost no change in the line width
\citep{coll95,cohen05}.  As a result, the observational parameter $v\sin i$
may systematically underestimate the true projected equatorial rotation
velocity \citep{town04}. In contrast,  \citet{cran05} found that this may
be true for the late-type Be stars (i.e., spectral types B3 and later), but
early-type Be stars do seem to be consistent with a range of intrinsic
rotation speeds between $\sim$40\% and 100\% critical. For stars rotating
sufficiently below the critical velocity, any physical model for the origin
of Be-star discs requires a significant increase in angular momentum above
the photosphere.

While the origin of the gas in these discs is agreed to be material ejected
from the stellar photosphere, the precise mechanism triggering this mass
ejection remains elusive.  One of the first models that tried to explain
the formation of circumstellar discs is the wind-compressed model 
\citep{bjor93}.  The fundamental idea behind this model is that in a
rotating wind, the material in the wind tends to orbit the star during the
time it is accelerated outward. If the outward acceleration is small
compared to the rotation, then the radiation-driven wind flows toward the
equator. Complimentary streamlines from opposite hemispheres would
intercept each other at the equatorial plane giving rise to a pair of
shocks above and below the equator and eventually the disc. Note that the
disc is not a result of enhanced equatorial mass-loss. A major problem of
this model arises when  non-radial components are included in the radiation
force, as these components lead to an effective suppression of the
equatorward flow needed to compress the wind \citep{owoc96}. In addition,
this model was unable to reproduce the IR excess of Be stars
\citep{porter97}.

The magnetically torqued disc model \citep{brown08} extends the ideas of
the wind compressed disc model by adding magnetic steering and
torquing up of the angular momentum of the wind. The up-flowing wind from
near the star is magnetically channeled back to the equator. This obviates
the \citet{owoc96} objections such as the driving of a polar flow by the
non-radial radiation field. The magnetically torqued disc model requires a
dipole-like magnetic field to channel the flow of wind material into a disc
region along the equatorial plane.

An alternative scenario that has been put forward to elevate disc matter to
orbits well above the stellar surface and which is increasing in popularity
is non-radial pulsations \citep[][and references therein]{cran09}.  
Non-radial pulsations may give rise to outwardly propagating circumstellar
waves, which inject enough angular momentum into the upper atmosphere to
spin up a Keplerian disc. The main uncertainty in this model is the
connection between pulsations and resonant waves since pulsations
themselves are evanescent in the stellar photosphere. However, once the
resonant waves have been formed, they grow in amplitude with increasing
height, begin to propagate upwards, and steepen into shocks. The resulting
dissipation would create substantial wave pressure that both increases the
atmospheric scale height and transports angular momentum upwards.

Once the disc is formed, the viscous decretion disc model
\citep{lee91,okazaki01} stands up as a very promising model that explains
many of the observed features in discs around Be stars. In this model,
matter supplied from the equatorial surface of the star gradually drifts
outward by the viscous torque and forms the disc. Decretion discs operate
in a similar fashion to accretion discs, except that the sign of $\dot{M}$
(mass decretion/accretion rate) is opposite. Viscosity is treated following
the Shakura-Sunyaev's $\alpha$-prescription. The outflow is highly subsonic
in the inner part of the disc and close to Keplerian. Although aspects of
the modeling (e.g., the supply of matter with sufficient angular
momentum to the inner edge of the disc or the interaction of the stellar
radiation with the disc matter) are not fully solved yet, the model is able
to explain the very low outflow velocities (the observed upper limits are,
at most, a few km s$^{-1}$) and also the V/R variability observed in Be
stars.

\begin{table*}
\begin{center}
\caption{Disc-loss episodes in Be/X-ray binaries.}
\label{loss}
\begin{tabular}{lccccccc}
\tableline
Source		&Spectral 	&$V_0$	&$(B-V)_0$&EW(H$\alpha$)&$T_{\rm disc}$ &$T_{\rm loss}$ &Reference\\
name		&type		&	&	&	(\AA)	&(year)		&(day)		&\\
\tableline
4U 0115+63	&B0.2Ve		&15.5	&1.43	&+2.4	&3-5	&800	&1,2\\
X Per		&B0Ve  		&6.80	&0.14	&+2.7  	&7	&800	&3,4\\
1A 0535+262	&O9.7IIIe	&9.44	&0.41	&+2.5	&4-5	&600	&5,6\\
RX J0812.4--3114&B0.2IVe 	&12.5	&0.4	&+1.4	&4	&--	&7\\   	  
RX J0440.9+4431	&B0.2Ve		&--	&--	&--0.5	&$>$10	&$<$200	&8\\
IGR J06074+2205	&B0.5Ve		&--	&--	&+2.2	&$>$5	&--	&9 \\
SAX J2103.5+4545&B0Ve		&14.35	&1.07	&+2.8	&1.5-2	&1000	&10\\
\tableline
\end{tabular}
\tablecomments{[1] \citet{negu01}, [2] \citet{reig07}, [3] \citet{roch97}, [4] \citet{clar01}
[5] \citet{haig99}, [6] \citet{clar98}, [7] \citet{reig01}, [8] \citet{reig05}
[9] \citet{reig10b}, [10] \citet{reig10a}}
\end{center}
\end{table*}

\subsection{Disc-loss episodes}

The use of photometric calibration for the determination of the spectral
type and luminosity class in a Be star is not as straightforward as in a
non-emission-line B-type star owing to the presence of the surrounding
envelope, which distorts the characteristic photospheric spectrum.  Be
stars appear redder than the non-emission B stars, owing to the additional
reddening caused by the hydrogen free-bound and free-free
processes in the circumstellar envelope. 

Thus, in a Be star, one has to correct for both circumstellar and
interstellar reddening before any calibration can be used. There is no easy
way to decouple these two reddening contributions. Although some iterative
procedure has been proposed to correct for both circumstellar and
interstellar reddening from narrow-band photometry \citep{fabr98} or using
infrared colours \citep{howells01,dougherty94}, the most reliable
measurement of the true photometric colours of a Be star is when the disc
contribution from the disc is negligible. That is the reason that disc-loss
episodes are so important.  The main accepted disc-loss indicator is the
\ha\ line (Fig.~\ref{be}). When the star loses the disc, the H$\alpha$ line
shows an absorption profile and the X-ray activity ceases or is largely
reduced.

Table~\ref{loss} list the \bex\ that have gone through disc-loss phases. It
also shows the typical time scales associated with disc variability and the
intrinsic colours of the underlying B star (not corrected for interstellar
reddening). $T_{\rm disc}$ is the typical duration of formation/dissipation
of the circumstellar disc and  $T_{\rm loss}$ represents the maximum
duration of the disc-loss phase.

\section{Observational evidence of the disc-neutron star interaction}

Be stars may also exist as single objects, i.e., not forming part of an
X-ray binary system. There is general consensus that the optical/IR long-term
(months to years) variability seen in Be stars is caused by the equatorial
disc around the massive star. In \bex, the X-ray variability is attributed
to changes in the accretion rate. Since the circumstellar disc constitutes
the fuel that powers the X-ray machinery, the X-ray variability can also be
ascribed to changes in the disc. The Be star in a \bex\ is
assumed to have the same physical properties (mass, radius, luminosity) as
isolated Be stars. Therefore, any difference in the variability patterns
of isolated and \bex\ must be the result of different structure
and physical properties of the disc (size, density). 

The following questions arise: Are the equatorial discs of isolated Be
stars different from those of Be stars forming part of an X-ray binary? If
yes, what causes this difference?

The obvious difference between an isolated Be star and a \bex\ is the
presence of a neutron star in the latter. Since the underlying Be star has
the same physical characteristics in both type of systems, we conclude that
any difference in the properties of the disc in binaries must be attributed
to the neutron star.

Traditionally, the interaction between the two components in a \bex\ was
thought to go into one direction, namely, from the massive companion to the
neutron star, through the transfer of mass. Matter from the Be star
captured by the neutron star is accreted and powers the X-ray emission.
Given the large mass ratio of the two components it was believed that the
neutron star exerted little effect on the massive companion. This picture is
now changing, as new observational evidence for the interaction
on the neutron star and the circumstellar disc emerges.

In this section, I summarise the observational evidence gathered in recent
years of how the compact object affects the optical/IR emission properties
of a \bex. The line of reasoning is the following: \bex\ and single Be show
different optical/IR patterns of variability because the equatorial disc
has different physical properties. The disc exhibits different properties
because of the presence or absence of an external agent, that is, the
neutron star.  The neutron star controls somehow the evolution of the disc
because it prevents its free expansion and causes truncation.

In searching for observational evidence for the effects that the neutron star
produces on the circumstellar disc of the Be star in \bex, I have followed
two different approaches

\begin{enumerate}

\item {\it Correlations with the orbital period}. Since \bex\ are in general
moderately eccentric systems, any property that correlates with the orbital
period of the system can be interpreted as due to the action of the neutron
star. The logic behind this idea is that the shorter the orbital period,
the smaller the orbit. Hence circumstellar discs in narrow-obit systems must
be more affected by the gravitational pull exerted by the neutron star
during each periastron passage than systems with longer orbital periods.

\item {\it Comparison of the variability patterns in isolated Be stars and
\bex}. If there is no physical difference between the B star in an isolated
Be and the B star in a \bex, and the optical/IR variability is driven by
changes in the disc, then any difference in the observational properties of
these two types of systems must come from the effect of the neutron star on
the disc. The objective here is to search for such differences in the
evolutionary patterns of various physical parameters (disc density and
radius, the strength and shape of spectral lines, particularly
H$\alpha$, and/or characteristic time scales).

\end{enumerate}

This section begins with a summary of the viscous decretion disc model (see
also Sect.~\ref{discform}) and it is followed by a compilation of the
observational evidence of the interaction of the neutron star with the
circumstellar envelope of the Be star.

\begin{figure}[t]
\centering
\includegraphics[width=0.95\linewidth]{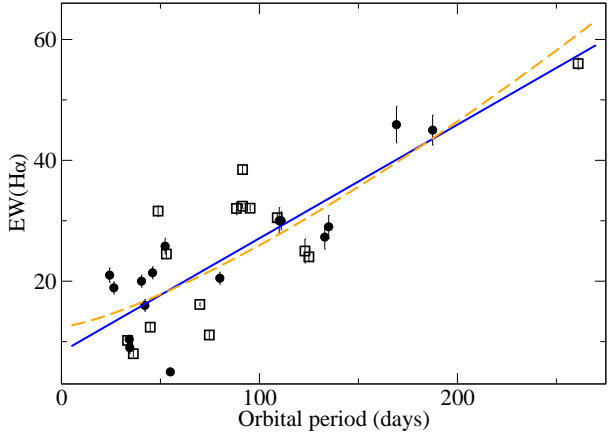}
\caption{P$_{\rm orb}$--EW(H$\alpha$) diagram. Milky way objects have
been represented by circles and \bex\ in the Small Magellanic Clouds by 
squares. The linear regression for the combined data sample is shown with a 
solid line, while the dashed line represents a power-law fit with exponent
n=4/3.}
\label{ewpo}
\end{figure}

\subsection{Disc truncation}

The idea of disc truncation in \bex\ was first suggested by \citet{reig97}
from an observational point of view and the theory subsequently developed
by Okazaki and collaborators in a series of papers:
\citet{neguoka01,okaz01,okaz02}.

The central idea behind the viscous decretion disc models is that Be stars'
discs are supported by viscosity. In this respect, the viscous excretion
discs are very similar to the well-known viscous accretion discs, except
for the changed sign of the rate of the mass flow. Angular momentum is
transferred from the optical companion to the inner edge of the disc,
increasing its angular velocity to Keplerian (rotationally supported). The
radial velocity component is subsonic even at the distance where the
neutron star lies.  Truncation occurs by the tidal interaction when the
resonant torque exerted by the neutron star exceeds the viscous torque.
This occurs only at certain radii --- where the ratio between the angular
frequency of disc rotation and the angular frequency of the mean binary
motion is a rational number. The efficiency of truncation depends strongly
on the gap, $\delta r$, between the truncation radius and the inner
Lagrangian point and the viscosity parameter $\alpha$. If $\tau_{\rm
drift}$ is the time scale for a particle in the disc to cross this gap then
truncation will be efficient when $\tau_{\rm drift} > P_{\rm orb}$. Typical
values of the viscosity parameter are $\alpha < 1$. Truncation is expected
to be more efficient in low and moderate eccentricity systems ($e\simless
0.3$) with narrow orbits ($P_{\rm orb} \simless 40$ d) than in high
eccentricity systems because the former have wider gaps. 

\citet{okaz02} performed numerical simulations of the disc formation around
isolated Be stars and of the interaction between the Be-star disc and the
neutron star in \bex\ and showed that resonant truncation is effective if
the viscosity parameter is $\alpha << 1$. There is a radius outside of
which the  azimuthally-averaged surface density decreases steeply.  When
$\alpha \sim 0.1$ the slope outside this radius is steeper,  giving a
stronger truncation effect on the disc. In the truncated disc model, the
material lost from the Be star accumulates and the disc becomes denser
more rapidly than if around an isolated Be star.

\subsection{Correlations with orbital period}

If the circumstellar disc is responsible for the variability observed at
all wavelengths in \bex, then correlations of an observational parameter
with the orbital period can be attributed to the interaction of the neutron
star with the disc.

\subsubsection{The P$_{\rm orb}$--EW(H$\alpha$) diagram}
\label{ewprob}

Figure \ref{ewpo} shows an updated version of the P$_{\rm
orb}$--EW(H$\alpha$) diagram by \citet[][see also
\citet{reig07b,anto09b}]{reig97}. The original P$_{\rm orb}$--EW(H$\alpha$)
diagram contained 11 \bex, of which only 9 had well established orbital
periods. Since then several new \bex\ have been discovered and their
orbital periods determined. Especially remarkable has been the discovery of
a large number of \bex\ in the Magellanic Clouds
\citep{coe05,liu05,schm06,anto09b}. In this figure, only sources with
well-determined orbital periods have been included, thus other SMC X-ray
pulsars (such as SXP756, SXP8.80, SXP46.6, and SXP304) with published
H$\alpha$ equivalent widths are not present in the figure.

The P$_{\rm orb}$--EW(H$\alpha$) diagram is based on the fact that the
H$\alpha$ line is the prime indicator of the circumstellar disc state.
Although an instantaneous measurement of the EW(H$\alpha$) may not be an
effective measurement of the size of the disc, the maximum equivalent
width, when monitored during a long length of time (longer than the typical
time scales for changes in the circumstellar disc, namely, from  few months
to a few years) becomes a significant indicator of the size of the disc. 
Based on interferometric observations, \citet{quir97} and \citet{tycn05}
have shown that there is a clear linear correlation between the net
H$\alpha$ emission and the physical extent of the emitting region. Also,
the observed correlations between the spectral parameters of the H$\alpha$
line (FWHM, EW, peak separation in double-peak profiles) and the rotational
velocity that have been observed in many Be stars are interpreted as
evidence for rotationally dominated circumstellar disc \citep{dach86}. In
particular, interpreting the peak separation ($\Delta_{\rm peak}$) of the
H$\alpha$ split profiles as the outer radius ($R_{ \rm out}$) of the
emission line forming region \citep{huan72}

\begin{equation}
\frac{R_{ \rm out}}{R_*}=\left(\frac{2 v \sin i}{\Delta_{\rm peak}}\right)^2 
\label{huang}
\end{equation}

\noindent the radius of the emitting region can be estimated
\citep{humm95,jasc04}. $v\sin i$ is the projected rotational velocity of
the B star ($v$ is the equatorial rotational velocity and $i$ the
inclination toward the observer). As the EW(H$\alpha$) increases, the peak
separation decreases, hence increasing the radius of the H$\alpha$ emitting
region \citep{dach92,hanu88}. Single peak profiles typically correspond to
larger discs. 

Note that the value of the H$\alpha$ equivalent width of Fig.~\ref{ewpo} is
the maximum ever reported. Some systems, especially in the SMC systems,
have been discovered recently, hence the monitoring of the H$\alpha$ is
necessarily short. For these systems, the value of the EW(H$\alpha$) may not
represent a high state of the circumstellar disc. The EW(H$\alpha$) of such
systems should be considered as lower limits. This is the case of the
galactic \bex\ GRO J2058+42 and most likely of the SMC source SXP455 (RX
J0101.3-7211). 

Assuming then, that EW(H$\alpha$) provides a good measure of the size of
the circumstellar disc, Fig.~\ref{ewpo} indicates that systems with long
orbital periods have larger discs, while narrow orbit systems contain
smaller discs. The existence of the above relationship 
provides a  strong argument that the discs are
truncated, due to the action of the compact object, to a radius  that is a
fraction of the semimajor axis.  Given the  similarity in masses of the
optical companions in \bex, we would expect  that the truncation radius
$r_{\rm trunc}$ would scale with the semimajor  axis according to Kepler's
Third Law as $r_{\rm trunc} \propto P^{2/3}$.  Since \ew\ scales as the
square of the disc radius  all other things being equal
\citep{grundstrom06},  the simple prediction would be $EW(H\alpha) \propto
P^{4/3}$. The solid line in Fig.~\ref{ewpo} corresponds to a simple linear
regression fit, while the dashed line represents a power-law fit with
exponent equal to 4/3.

\begin{figure*}[t]
\includegraphics[width=0.95\linewidth]{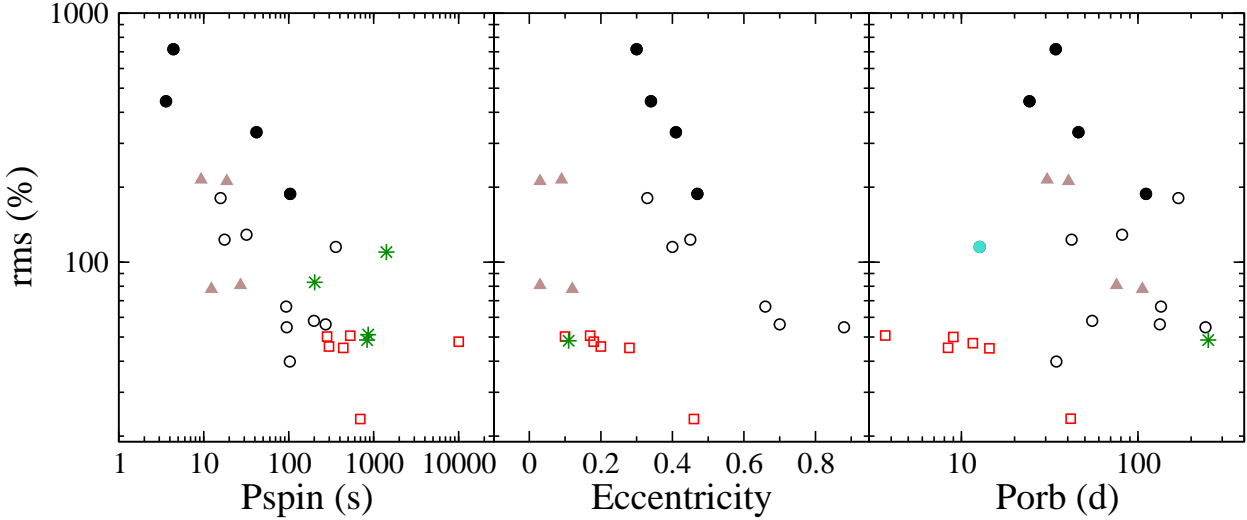}
\caption[]{X-ray $rms$ variability of the {\it RXTE}/ASM light curves as a function of the
spin period (left), eccentricity (middle) and orbital period (right) of
galactic \bex. Black filled circles denote systems that have shown type II 
outbursts.
Squares represent SGXBs and the star symbols correspond to persistent \bex.
The cyan filled circle on the right panel correspond to SAX J2103.5+4545 
(see Sect.~\ref{intro_hmxb}). From \citet{reig07b}.}
\label{rmsorb}
\end{figure*}

\subsubsection{Be $\longrightarrow$ B $\longrightarrow$ Be phase}

Column 4 of Table~\ref{loss} gives the duration of the disc
formation/dissipation phase, $T_{\rm disc}$, of those \bex\ for which 
reliable data exist. The orbital period is given in column 6. As can be
seen there is a good correlation between $T_{\rm disc}$ and the orbital
period. Systems with short orbital periods tend to show faster disc growth
and dissipation cycles, while longer time scales are associated with longer
orbital periods. This result agrees with the scenario that the Be disc is
truncated by the tidal torques from the neutron star, which would be
more significant for narrower orbits.

\subsubsection{Long-term X-ray variability}

\citet{reig07b} investigated the long-term X-ray variability, expressed as
the root-mean-square ($rms$) of the ASM/{\it RXTE} light curves, of a set
of \bex\ and found that the systems with the larger $rms$ are those
harbouring faster rotating neutron stars and those in low eccentric and
relatively narrow orbits. The $rms$ amplitude was computed as
$rms=\sigma^2/\bar{x}^2$, where $\bar{x}$ is the mean count rate and
$\sigma^2=\sigma_{\rm obs}^2-\sigma_{\rm exp}^2$ is the difference between
the observed variance, $\sigma_{\rm obs}^2=\sum_{i} (x_i-\bar{x})^2/N$, and
the expected variance, $\sigma_{\rm exp}^2=\sum_{i} \sigma_i^2/N$
($\sigma_i$ are the experimental errors, and $N$ is the total number of
points).

Figure \ref{rmsorb} shows the $rms$ as a function of the spin period,
eccentricity and orbital period of the systems. Note that by using the
$rms$ the emphasis is put on the amplitude of the variations rather than on
how often the variations take place. For the sake of comparison, persistent
\bex\ and supergiant X-ray binaries, that is,  systems whose optical
companion is an evolved (luminosity class I-II) star, have also been
included. Different type of systems have been represented by different
symbols as follows: transient \bex\ by circles, persistent \bex\ by stars
and supergiants by squares. Systems that have shown Type II outbursts are
represented by black filled circles. Triangles denote the class of low-$e$
\bex\ (Sect.~\ref{low-e}).  It is very likely that low-$e$ \bex\ and
persistent \bex\ form  only one class of systems. In fact,  X-Per, which is
considered the prototype of the persistent \bex, also belongs to the class
of low-$e$ \bex. 

In all cases a clear anticorrelation is apparent: the systems with the
faster rotating neutron stars and with low eccentricity and narrow orbits are
more variable, i.e., present higher $rms$. This plot then suggests two
interesting results. First, since the \bex\ and the SGXB occupy clearly
distinct regions in Fig.~\ref{rmsorb}, the X-ray variability in high-mass
X-ray binaries does not only depend on the physical conditions in the
vicinity of the compact object but also on the mass transfer
mechanism, i.e., whether a circumstellar disc is present or not. Second, the
systems containing fast spinning neutron stars are more likely to exhibit Type II
outbursts.

In general, {\em i)} the more eccentric the orbit the lower the variability
and {\em ii)} type II activity mainly occurs in low-eccentric systems.  The
low eccentricity class of \bex\ follow the general trend in the $rms-P_{\rm
spin}$ and $rms-P_{\rm orb}$ plots but clearly distinguish themselves from
the classical \bex, in the $rms-e$ diagram.  The anticorrelation of the
$rms$ with the orbital period provides further evidence for the truncation
of the disc in systems with short orbital periods, i.e., those having
narrow orbits.

A prediction of the viscous decretion disc model is  that type I
outbursts are expected to occur more often in high-eccentric systems,
whereas type II outburst would dominate in low-eccentric systems, in
agreement with the results shown in Fig.~\ref{rmsorb}.

\citet{coe05} carried out a major study of the optical and infrared
characteristics of \bex\ in the SMC. They found that the optical photometric
variability is greatest when the circumstellar disc size is least and also
in systems containing fast rotating neutron stars. \citet{reig07b} extended
these results to the X-ray domain. While the works by \citet{okaz01},
\citet{zhan04a} and the $EW(H\alpha)-P_{\rm orb}$ correlation clearly
identify truncation with low eccentric and narrow orbit systems, the plots
of the {\em X-ray} $rms$ as a function of the system parameters
(Fig.~\ref{rmsorb}) and those of the  EW(H$\alpha$) and $P_{\rm spin}$ as a
function of the {\em optical} photometric $rms$ \citep[Figs 6 and 7 in
][]{coe05}  identify truncation with variability. That is, systems in which
truncation is favoured are the most variable ones in both, the X-ray and
optical bands.

\subsection{Be in \bex\ {\it versus} isolated Be}

The other line of argument to demonstrate that the neutron star exerts a
measurable effect on the optical companion is based on the comparison of
the variability patterns of the optical/IR emission in isolated Be stars
and \bex.

\subsubsection{V/R variability}

As explained in Sect.~\ref{vr}, V/R variability is defined as the intensity
variations of the two peaks (known as violet and red peak) in the split
profile of a spectral line. In many Be stars, if monitored over a long
enough period of time, these variations are quasiperiodic \citep{okaz97}. 
V/R variability is currently explained by the global one-armed oscillation
model. This model  suggests that the  long-term  V/R  variations  are
caused by global $m=1$ oscillations in the cool equatorial disc of the Be
star.  In other words, an enhanced density perturbation develops on one
side of the disc, which  slowly  precesses.  The  precession  time is
that associated with the V/R  quasi-period.   The periods of V/R variations
in isolated Be stars range from years to decades, with a statistical mean
of 7 years, which is much longer than the rotational period of the central
star \citep{okaz97}.  In contrast, in \bex\ (Table~\ref{timescales}), the V/R
quasiperiods, $T_{\rm V/R}$, are significantly shorter ($\simless 5$ yr ),
which can be interpreted as the result of smaller discs in \bex\ due to
truncation caused by the neutron star. 

\citet{oktariani09} studied the tidal effect of the companion on the global
oscillation modes in equatorial discs around binary Be stars. They found
that the oscillation modes are well confined when the disc is larger than a
few tens of stellar radii.  In smaller discs, however, the mode confinement
is incomplete and the oscillation period depends on the binary parameters
significantly. The oscillation period becomes longer in the case of a wider
binary separation and/or a lower binary mass ratio. Such dependence is
observed only for short period binaries. In wide binaries where the disc
size is large enough to confine the one-armed modes in the inner part of
the disc, the mode characteristics depend little on the binary parameters.

\begin{table*}
\begin{center}
\caption{Characteristic timescales for disc variability in Be/X-ray binaries.}
\label{timescales}
\begin{tabular}{llccccl}
\tableline
X-ray	&Optical&Spectral &$T_{\rm disc}$ &$T_{\rm V/R}$ &$P_{\rm orb}$ &Reference\\
name	&name	&type	&(year)		&(year)		&(day)		&\\
\tableline
4U 0115+63	&V635 Cas	&B0.2Ve		& 3-5	&0.5-1.5	&24.3	 &1\\
V 0332+53	&BQ Cam		&O9Ve		& 4-5	&1		&34.2	 &2,3\\
4U 0352+309	&X Per		&B0Ve		& 7	&0.6-2		&250	 &4\\
1A 0535+262	&V725 Tau	&O9.7IIIe	& 4-5	&1-1.5		&111	 &5,6\\
RX J0812.4-3114	&LS 992		&B0.2IVe	& 4	&$-$		&80	 &7\\		
SAX J2103.5+4545&--		&B0.5V		&1.5-2	&$-$		&12.7	 &8 \\
RX J 0146.9+6121&LS I +61 235	&B1Ve		&$>$10	&3.4		&$>$200\tablenotemark{*}&9\\
4U 1145-619	&V801 Cen	&B1Ve		&$>$10	&3		&186	 &10\\
RX J0440.9+4431	&LS V +44 17	&B0.2Ve		&$>$10	&$-$		&$>$150\tablenotemark{*}&11\\
IGR J06074+2205 &--		&B0.5V		&$>$5	&$\sim$5	&--	 &12 \\
\tableline
\end{tabular}
\tablenotetext{*}{Obtained from the $P_{\rm spin}-P_{\rm orb}$ correlation.}
\tablecomments{[1] \citet{negu01}, [2] \citet{negu99}, [3] \citet{gora01}
[4] \citet{clar01}, [5] \citet{haig04}, [6] \citet{clar98}
[7] \citet{reig01}, [8] \citet{reig10a}, [9] \citet{reig00}
[10] \citet{stev97}, [11] \citet{reig05}, [12] \citet{reig10b}}
\end{center}
\end{table*}
\begin{figure}[t]
\centering
\includegraphics[width=0.9\linewidth]{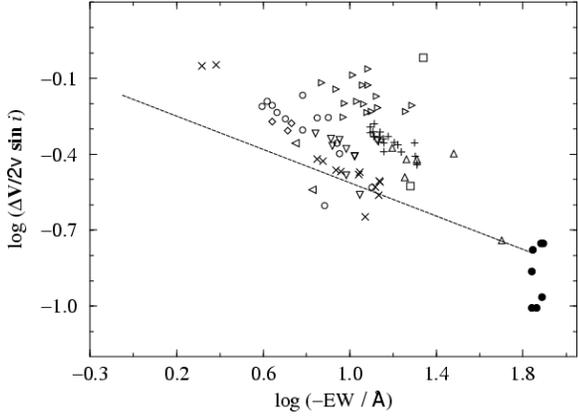}
\caption{Peak separation as a function of \ew. The dashed line represents 
the average behaviour of the isolated Be stars. 
Be/X-ray binaries show, on average denser disks. Different sysmbols
correspond to different \bex. From \citet{zama01}.
\label{dense}}
\end{figure}

\subsubsection{Disc density}

\citet{zama01} carried out a comparative study of the circumstellar discs
in \bex\ and isolated Be stars based on the \ha\ line emission properties.
They found that the discs of \bex\ are twice as dense as and have, on
average, smaller size than the discs of isolated Be stars.

\citet{hanu88,hanu89} showed that there is a correlation between the \ha\
peak separation and the equivalent width in Be stars, according to the law

\begin{displaymath}
\label{den}
\log \left( \frac{\Delta V} {2\,v\,\sin{i}} \;\right) = 
 a \log \left( \frac {-EW({\rm H}\alpha)}{\rm\AA}\;\right) +\;b 
\end{displaymath} 

\noindent where $v \sin i$ is the projected rotational velocity and
EW(H$\alpha$) is given in Angstroms. $a$ and $b$  are related to the
rotational law index $a=-j/2$ ($j$=0.5 for Keplerian rotation and $j$=1 for
conservation of angular momentum) and with the disc electron density,
respectively.  Figure~\ref{dense} shows the plot of $\log \frac{\Delta V}
{2\,v\,\sin{i}}$ {\em vs} $\log {\rm EW(H\alpha)}$ for a number of \bex.
The straight line ($a=-0.4$, $b=-0.1$) represents the average behaviour of
the isolated Be stars investigated by \citet{hanu89}. The data points
correspond to \bex. As can be seen, the vast majority of the data points
lie above the average line, i.e. they are shifted towards denser discs. 

Another result that supports denser disc in Be binaries is the correlation
observed between the H$\alpha$ equivalent width and the $(B-V)$ colour
excess in a sample of Be/X-ray binaries in the Magellanic Clouds
\citep{torr07}.  The total reddening in the direction of a Be star or \bex\
binary is composed of two components: one (constant) produced mainly by
dust in the interstellar space through the line of sight $E^{is}(B-V)$ and
another one (variable) produced by the circumstellar gas around the Be star
$E^{cs}(B-V)$. The nature and wavelength dependence of these two reddenings
is completely different but their final effect upon the photometric indices
and spectrum is, in principle, indistinguishable. \citet{torr07} found that
the circumstellar excess is $\sim$ 10 times larger for Be stars in binaries
than for isolated Be stars with the same amount of emission, and also that
the circumstellar colour excess, $E^{cs}(B-V)$ saturates at$\sim$ 0.2
magnitudes beyond EW(H$\alpha$)$\approx$20 \AA.

Again the denser discs seen in \bex\ can be explained by disc truncation.
\citet{okaz02} showed that during the initial build-up phase, the disc
evolution is similar to that for isolated Be stars. But, later on, the
effect of the resonant torque becomes apparent, preventing the disc gas
from drifting outwards at several resonance radii. As a result, the disc
density increases more rapidly than that for isolated Be stars, in
agreement with the results presented in this section.

\subsubsection{Disc size}

As mentioned above, interferometric observations have shown that there is a
clear linear correlation between the net H$\alpha$ emission and the
physical extent of the emitting region \citep{quir97,tycn05}. Using
numerical models of the circumstellar discs of Be stars,
\citet{grundstrom06} showed that there are monotonic relationships between
the emission-line equivalent width and the ratio of the angular half-width
at half maximum of the projected disc major axis to the radius of the star,
providing a method to estimate disc radii based only on the H$\alpha$
equivalent width with an accuracy of about 30\%. Since, on average,
isolated Be stars show larger \ew\ than Be stars in \bex\
\citep{reig97,anto09b}, we conclude that, on average, the envelopes of Be
stars in \bex\ are smaller than those present in isolated Be stars.

\subsection{Quantized infrared excess flux states}

Further evidence in favour of the decretion disc model and disc truncation
comes from the infrared band. \citet{haig04} showed that the Be/X-ray
binary A 0535+26 displays a clear bimodality in the infrared magnitudes
distribution, suggesting that the system alternates between a faint and a
bright flux state.

These quantized infrared excess flux states are a natural consequence of
the resonant truncation hypothesis proposed by \citet{okaz01} and
constitute a firm observational verification of the proposed resonant
truncation scheme for Be X-ray binaries. According to this model, tidal
torques, which operate most strongly in orbits resonant with that of the
neutron star, truncate the disc at radii meeting the condition $P_{\rm
NS}:P_{\rm trunc}=n:1$ where n is an integer. \citet{haig04} identified 
the two states with resonances 5:1 and 6:1. A possible brighter infrared
state corresponding to resonance 4:1 may have also been seen during the
70's.

The simultaneity of X-ray activity with transitions between these states
strongly suggests a broad mechanism for outbursts, in which material lost
from the disc during the reduction of the truncation radius is accreted by
the neutron star.

The idea is simple and neat. When the disc outer radius makes a quantum
leap inwards the material present between those radii must be relocated
elsewhere in the system and becomes available for accretion. In 1993/94 a
change of infrared state as accompanied by a giant X-ray outburst. This
simultaneity of the type II outburst with the change of infrared state
strongly suggests that the giant outburst was caused by the accretion of
material previously resident between the 5:1 and 6:1 resonances.
Importantly, truncation would appear to be the cause of the X-radiation and
not vice versa, as truncation commenced before X-ray emission.

\section{Be/X in other galaxies}

The study of X-ray source populations in other galaxies is hampered by the
large distance from us. Even other galaxies of the Local Group are too
far to reach the quiescence population of HMXBs ($L_X \sim10^{33}-10^{35}$
erg s $^{-1}$). Good enough signal-to-noise optical  spectra, needed to
derive the spectral type of the optical counterparts, are very hard to
achieve. Because of the relatively low intergalactic extinction,
well-established distance and their proximity, the Magellanic Clouds (MC)
represent the best targets to study a complete X-ray binary population. In
fact, they are the only galaxies for which meaningful statistical studies
can be performed (current space detectors can detect MC X-ray sources down
to a few $10^{33}$ erg s $^{-1}$). Moreover, the MC show different chemical
compositions and are heavily interacting with the Milky Way, which affects
their star formation history. Therefore, the study of stellar populations
in the MC is particularly rewarding.   

Although past X-ray missions with imaging capabilities, such as {\em
Einstein} \citep{wang92}, {\em ROSAT} \citep{habe00}, and {\em ASCA}
\citep{yoko03} have surveyed the MC, the deep surveys of the current
missions {\em XMM-Newton} \citep{sasa03,eger08}, {\em CHANDRA}
\citep{mcgo08}, {\em RXTE} \citep{gala08} and {\em INTEGRAL}
\citep{gotz06,coe10} complemented with observations from ground-based
telescopes in the optical/IR band
\citep{schm06,schu07,mcbr08,anto09a,anto09b} are the ones giving new
insights into population synthesis models.

The Small Magellanic Cloud (SMC) hosts an unexpected large population of
HMXBs. Initial estimates of the number of HMXBs in the SMC were based on
the mass of the SMC relative to that of the Milky Way. The SMC is a few
percent of the mass of the Galaxy, in which about 70 X-ray pulsars are
known. Therefore one would expect to find only two or three systems.
However, there are about 50 HMXBs in the SMC. Curiously, only one (SMC X-1)
of these systems is not a Be/X, that is, it contains a supergiant
companion.

The reduced metallicity of the SMC ($\sim$ one fifth of solar) cannot
explain, by itself, the large number of HMXBs in this galaxy
\citep{dray06}. In addition to reduced metallicity, a large star formation
rate is needed to explain this excess of HMXBs \citep{grim03}. The star
formation rate/mass of the SMC is 150 times that of the Milky Way. The star
formation rate may have been enhanced as a result of tidal interactions
between the MC and the Galaxy. However, the most recent close approach of
the SMC and LMC was 200 Myr ago - much longer than the evolutionary
timescale of Be/X-ray binaries. This implies that either there has been a
significant delay between the encounter of the SMC and LMC and the onset of
star formation, or that subsequent waves of star formation have given rise
to these Be/X-ray binaries \citep{harr04}.

The main question that we wish to address in this section is whether the
population of \bex\ in the MC differs from that of the Milky Way. We focus
on the statistical distributions of spin periods, spectral type and \ha\
equivalent width of the optical counterparts.

\subsection{Spectral-type}

The spectral distribution of the \bex\ population in the Milky Way is
characterised by a sharp peak at B0 and a relatively narrow width, namely,
no \bex\ are found beyond type B2 and earlier than type O8.
\citet{negu02,anto09b} showed that \bex\ in the LMC follow the same
distribution as those in our Galaxy, while \citet{mcbr08} and
\citet{anto09b} showed that the spectral distribution of \bex\ in the SMC is
also consistent with that of the Milky Way. The only difference might be
the presence of a very small number of \bex\ with spectral type later than B2
in the SMC. However, given the very low number of these systems and the
uncertainty in the spectral classification, this difference is not
statistically significant.

\subsection{\ha\ equivalent width}

On average, the \ha\ equivalent width (\ew) of the SMC population of \bex\
is larger than that of the Milky Way. \citet{anto09b} found that the \ew\
distribution of SMC \bex\ peaks in the interval from --15 to --25 \AA,
while the \ew\  of Galactic \bex\ distribute more uniformly in the interval
from --30 to 0 \AA. Likewise, the number of sources with \ew\ $>$ 50 \AA\
is larger in the SMC. For the LMC systems the sample is too small to draw
any conclusions. This trend may just be a selection effect. The
discovery of \bex\ in the SMC results mainly from X-ray observations. Due
to their larger distance we tend to detect the brightest systems, hence
those with a large reservoir of accreting material, i.e., those with larger
discs.

\begin{figure}[t]
\centering
\includegraphics[width=0.9\linewidth]{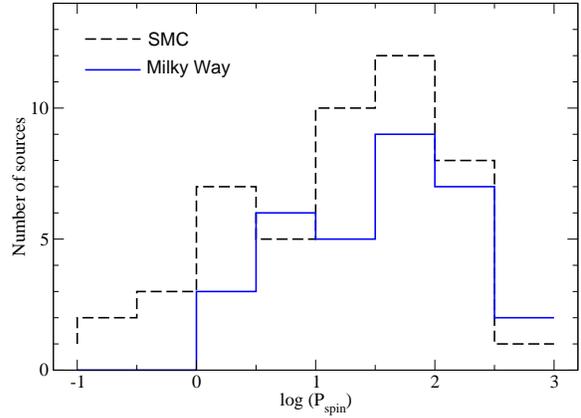}
\caption{Distribution of pulse periods in the SMC and the Galaxy.}
\label{spindist}
\end{figure}

\subsection{Spin periods}

As indicated in Sect.~\ref{stat}, virtually all HMXBs
are X-ray pulsars. Histograms of the spin period of HMXBs in the MC and
their comparison with Galactic sources can be found in \citet{layc05} and
\citet{liu05}. They both find that the SMC population appears to be shifted
to shorter periods. However, while this shift is very significant in the
\citet{layc05} sample, the difference is minor in the \citet{liu05} sample.
One contribution to the difference between the Galactic and SMC pulse
period distributions is the apparent lack of supergiant wind accretion
pulsars in the SMC. HMXBs with supergiant companions tend to have longer
($>100$ s) spin periods than \bex. As stated before only one such system is
known in the SMC. Figure~\ref{spindist} shows the spin period distributions
of well established \bex\ (i.e. excluding supergiant wind-powered systems) in
the SMC and Milky Way from the lists compiled by \citet{layc05} and
\citet{ragu05}. A Kolmogorov-Smirnov test gives no real evidence against
the null hypothesis that the two samples are drawn from the same
distribution.

\section{Summary}

The traditional picture of two classes of high-mass X-ray binaries, namely
supergiant X-ray binaries  and Be/X-ray binaries  is giving way to a more
complex situation where newly discovered systems may not fit in these
categories. Superfast X-ray transients, low eccentricity \bex, $\gamma$-ray
binaries, and $\gamma$-Cas like objects are among the new type of systems
that only recently have begun to emerge. \bex\ do not form a homogeneous
group either with transient and persistent sources, highly eccentric and
nearly circular orbits, and fast (P$_{spin}$ $\sim$ few seconds) and slow
(P$_{spin}$ $\sim$ few hundred of seconds)  rotating neutron stars.

Virtually, all \bex\ are X-ray pulsars. The study of the variability of the
X-ray spectral and pulse timing parameters across giant outbursts is
crucial to understand the physics of accretion in strongly magnetised
neutron stars. Only recently, attention is being paid to the aperiodic
variability and evolution of broad-band noise components across the
outburst. The application of diagrams such as hardness-intensity and
colour-colour diagrams, so widely used in black-hole and low-mass binaries,
to \bex\ and the definition of spectral states is a promising way to
understand the phenomenology of the evolution of outbursts in these
systems.

Progress in understanding \bex\ is also being made in the optical and
infrared bands where the massive companion shines bright. The data gathered
over the past ten years indicate that the interaction between the compact
object and the Be type star works in two directions: the massive companion
provides the source of matter for accretion; the amount of matter captured
and the way it is captured (transfer mechanism) change the physical
conditions of the neutron star (e.g. by spinning it up or down and shrinking
or expanding the magnetosphere). But also, the continuous revolution of the
neutron star around the optical counterpart produces observable effects,
the most important of which is the truncation of the Be star's equatorial
disc.

Future missions with improved sensitivities in the X-ray band will help
solve remaining open questions such as the formation of accretion discs
during X-ray outbursts, the origin of quasi-periodic oscillations,
characterisation of the population of persistent \bex\ and the triggering
mechanism of the giant outbursts.

\acknowledgments

I want to thank the referee, Douglas Gies, for his careful reading of this
manuscript and his useful and important comments.  This work has been 
supported  by the European Union Marie Curie grant MTKD-CT-2006-039965 and
EU FP7 "Capacities" GA No206469.

\nocite{*}
\bibliographystyle{spr-mp-nameyear-cnd}
\bibliography{biblio-u1}

\end{document}